\documentclass[Journal, draftclsnofoot, comsoc, onecolumn]{IEEEtran}

\ifCLASSINFOpdf

\else

\fi

\usepackage[nocompress]{cite}
\usepackage[dvipsnames]{xcolor}
\usepackage{color, soul}
\usepackage{graphicx}
\usepackage{amsmath}
\usepackage{epsfig}
\usepackage{subfigure}
\usepackage[ruled,linesnumbered]{algorithm2e}
\usepackage{subfigure}
\graphicspath{{eps/}}

\usepackage{amssymb}
\usepackage{amsthm}

\newcommand{\mynote}[1]{\definecolor{note}{RGB}{0,128,255}  \sethlcolor{note}\hl{\textbf{#1}}}

\DeclareMathOperator*{\argmax}{\arg\!\max}

\begin{document}

%





\title{MUVINE: Multi-stage Virtual Network Embedding in Cloud Data Centers using Reinforcement Learning based Predictions}



%

    \author{Hiren~Kumar~Thakkar, ~\IEEEmembership{Member,~IEEE}, Chinmaya~Kumar~Dehury, Prasan~Kumar~Sahoo,~\IEEEmembership{Senior Member,~IEEE}  
    \thanks{Hiren Kumar Thakkar was with the department of Computer Science and Information Engineering, Chang Gung University, Guishan, Taiwan. Currently, he is with the department of Computer Science Engineering, Bennett University, India. Email: hirenkumar.thakkar@bennett.edu.in.}
    \thanks{Chinmaya Kumar Dehury was with the department of Computer Science and Information Engineering, Chang Gung University, Guishan, Taiwan. Currently, he is with the Mobile \& Cloud Lab, Institute of Computer
Science, University of Tartu, Estonia. Email:
chinmaya.dehury@ut.ee.}
    \thanks{Prasan Kumar Sahoo (Corresponding author) is with the department of Computer Science and Information Engineering, Chang Gung University, Guishan, Taiwan. He is also an Adjunct Research Fellow in the Division of Colon and Rectal Surgery, Chang Gung Memorial Hospital, Linkou, Taiwan. Email: pksahoo@mail.cgu.edu.tw}}



%

\maketitle
\begin{abstract}
The recent advances in virtualization technology have enabled the
sharing of computing and networking resources of cloud data
centers among multiple users. Virtual Network Embedding (VNE) is
highly important and is an integral part of the cloud resource
management. The lack of historical knowledge on cloud functioning
and inability to foresee the future resource demand are two
fundamental shortcomings of the traditional VNE approaches. The
consequence of those shortcomings is the inefficient embedding of
virtual resources on Substrate Nodes (SNs). On the contrary,
application of Artificial Intelligence (AI) in VNE is still in the premature stage and needs further
investigation. Considering the underlying complexity of VNE that
includes numerous parameters, intelligent solutions are required
to utilize the cloud resources efficiently via careful selection of appropriate SNs for the VNE. In this paper,
Reinforcement Learning based prediction model is designed for the
efficient Multi-stage Virtual Network Embedding (MUVINE) among the cloud
data centers. The proposed MUVINE scheme is extensively simulated and evaluated against the recent state-of-the-art schemes. The
simulation outcomes show that the proposed MUVINE scheme consistently outperforms over the existing schemes and provides the promising results.
\end{abstract}

\mynote{}

\IEEEpeerreviewmaketitle

\section{Introduction}\label{sec:intro}
The cloud computing is a recent advancement in virtualization technology that can dynamically provision infinite computing resources to the end users on pay as you use basis. The primary
concern of any Cloud Service Provider (CSP) is to ensure the
quality of cloud services to the end users with optimum computing
resources that translate into the maximum profit. Accordingly, various heuristics-based, nature-inspired learning-based, and Artificial Intelligence (AI) based approaches are proposed for the efficient cloud resource
management. The AI is one of the promising alternatives that potentially aid in efficient cloud resource management. AI is known as the machine demonstrated intelligence
and it is considered as a group of techniques designed to tackle the
application-specific problems such as prediction, classification,
recognition, visualization etc. Prominent AI techniques such as
Fuzzy Logic, Machine Learning (ML) and Deep Learning (DL) exist to
deal with various problems in every walk of our life. Among them,
ML is typically used to process the text and numeric dataset.
Generally, ML techniques are broadly classified into three
categories such as supervised, unsupervised, and reinforcement
learning with primary objective to let the machine extract,
analyze, train, and build the prediction models on the top of the
training dataset. In supervised learning, machine learns from the
supplied labels often called as the observed true outcomes;
whereas, in unsupervised learning, the machine learns without the
known outcomes. Contrary to the supervised and unsupervised
learning, the agent (machine) learns in real-time from the
environment to optimize the given problem in reinforcement
learning.

Although AI techniques help to build an efficient prediction model, it poses numerous challenges as well. The foremost challenge is to choose
the right AI technique for a given application, dataset type, size, and nature of
the problem. However, considering the robustness, the recent years have
witnessed a myriad growth of applications of AI gradually
replacing the traditional approaches with improved accuracy. The
home automation \cite{zhang2016optimal}, self-driving car
\cite{shalev2016safe}, cardiac signal processing \cite{thakkar2019towards}, automated
cloud resource management \cite{mijumbi2015neuro} are
considered few of the potential applications that benefit the
most by AI. Among the aforementioned applications, automated cloud resource management is considered as the challenging one due to
the underlying complexity.


On the other hand, cloud computing environment provides the
computing and storage resources to the users, which encourages the users to execute the complex applications and
stores the raw data without concerning much about the storage
space \cite{2018_adaptive}.
The computing and storage as virtual resources are provision to the end users by
implementing the virtualization technology atop the substrate
resources \cite{GHOBAEIARANI2018191}.
Different service models are used to provide the virtual resources
such as the Infrastructure as a Service (IaaS), Software as a
Service (SaaS) etc. In order to access the resources using IaaS
model, users need to send the request in the form of Virtual
Network (VN). Typically, each VN request may comprised of a number of Virtual Machines
(VMs) with the corresponding resource configuration, and the network
topology that defines the communication among the VMs
\cite{sahoo2018lvrm}.

On receiving the VN request, CSP creates the
required VMs onto a set of suitably interconnected Substrate Nodes (SNs) and establishes the communication links among the VMs. The aforementioned process is called
as a Virtual Network Embedding (VNE).
In general, VNE refers to the embedding of the VMs and virtual links in a sequential or parallel manner. However, the SN utilization is one of the major research issues that
needs to be addressed while embedding the VN.
In the past decade, different approaches are followed to efficiently carry
out the VNE process. In \cite{sahoo2018lvrm}, graph theory is used as a tool for the VNE. In \cite{sahoo2018lvrm}, the VN request and the substrate network are treaterd as graphs and the sub-graph from the substrate network equivalent to that of the VN request is obtained. In \cite{Wei2018Imperfect}, Hidden Markov Model is used to model the VNE problem.
Ant colony optimization \cite{gao2013multi} and Particle
swarm optimization \cite{song2019constructive} are few of the machine learning methods that are applied in the VNE process.

The machine learning approaches may use the historical information
of the incoming VN request, substrate network, and the usage
history.
The VN request may includes the resource demand of VMs.
Each VM is further associated with the priority value decided in
accordance with the Service Level Agreement (SLA), which indicates the importance of the
corresponding VM in resource allocation process. Upon embedding
the VN onto the SNs, the actual resource usage
information is obtained.
On the other hand, historical information regarding the SNs includes the amount of the resources allocated to the VN request, actual resource usage by the VMs, throughput
of each SN, addition and removal of the SNs at different time stamps. It is assumed that the CSP is
empowered with different software tools to collect such historical
information \cite{googleTrace}.
The historical information also includes the type of VN requests accepted and rejected under different cloud scenarios. Further, such labelled data are useful in the embedding process to provision the
suitable SNs for each VN request type.
The supervised ML algorithms such as linear regression, support
vector machine and Naive Bayes can be considered as the suitable
alternatives for the aforementioned data. However, it is essential
to design the ML algorithms, which explore the cloud environment and exploit the VM placement strategies for the real-time embedding of VMs on SNs.

\subsection{Motivation}
The substrate resource
utilization plays a major role in embedding the VNs as it has the
direct impact on the acceptance ratio and revenue of the CSP.
Here, the substrate resource refers to as the CPU, memory, and
network. The utilization refers to as the percentage of the
resources utilized by the VMs out of the allocated. Usually, the exact amount of resources is allocated to VMs and the virtual links to fulfill the resource demand of the requested VNs. However, our analysis
of the historical dataset \cite{googleTrace} reveals
that resources normally stay underutilized leading to the overall
ineffective resource utilization.

\begin{figure}[t]
    \centering
    \subfigure[VN requests and resource capacity of $SN_1$ and $SN_2$.]
    {
        \includegraphics[width=0.43\linewidth]{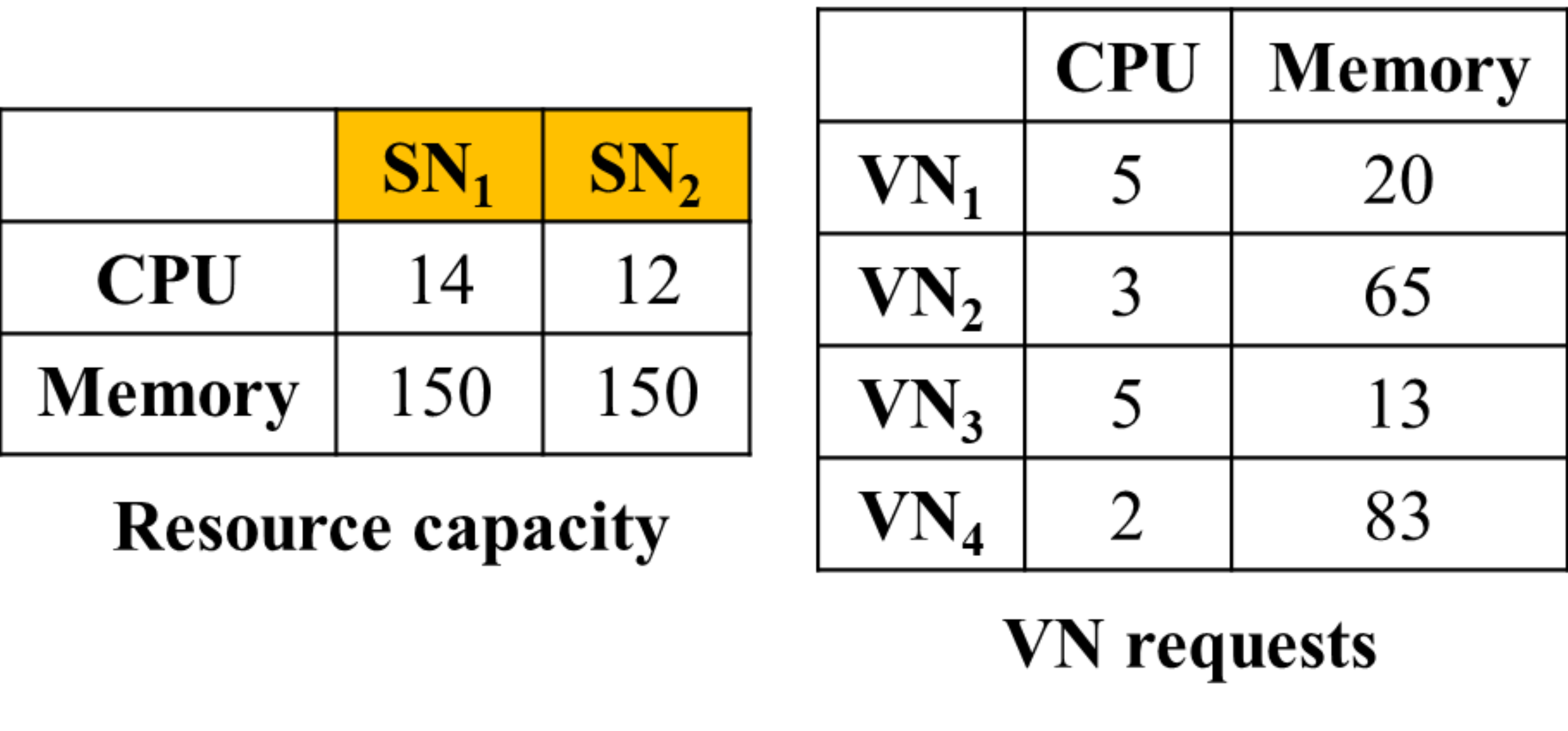}
        \label{fig:Motiv:Util_VN_req}
    }
    ~
    \subfigure[A new VN request $VN_5$.]
    {
        \includegraphics[width=0.23\linewidth]{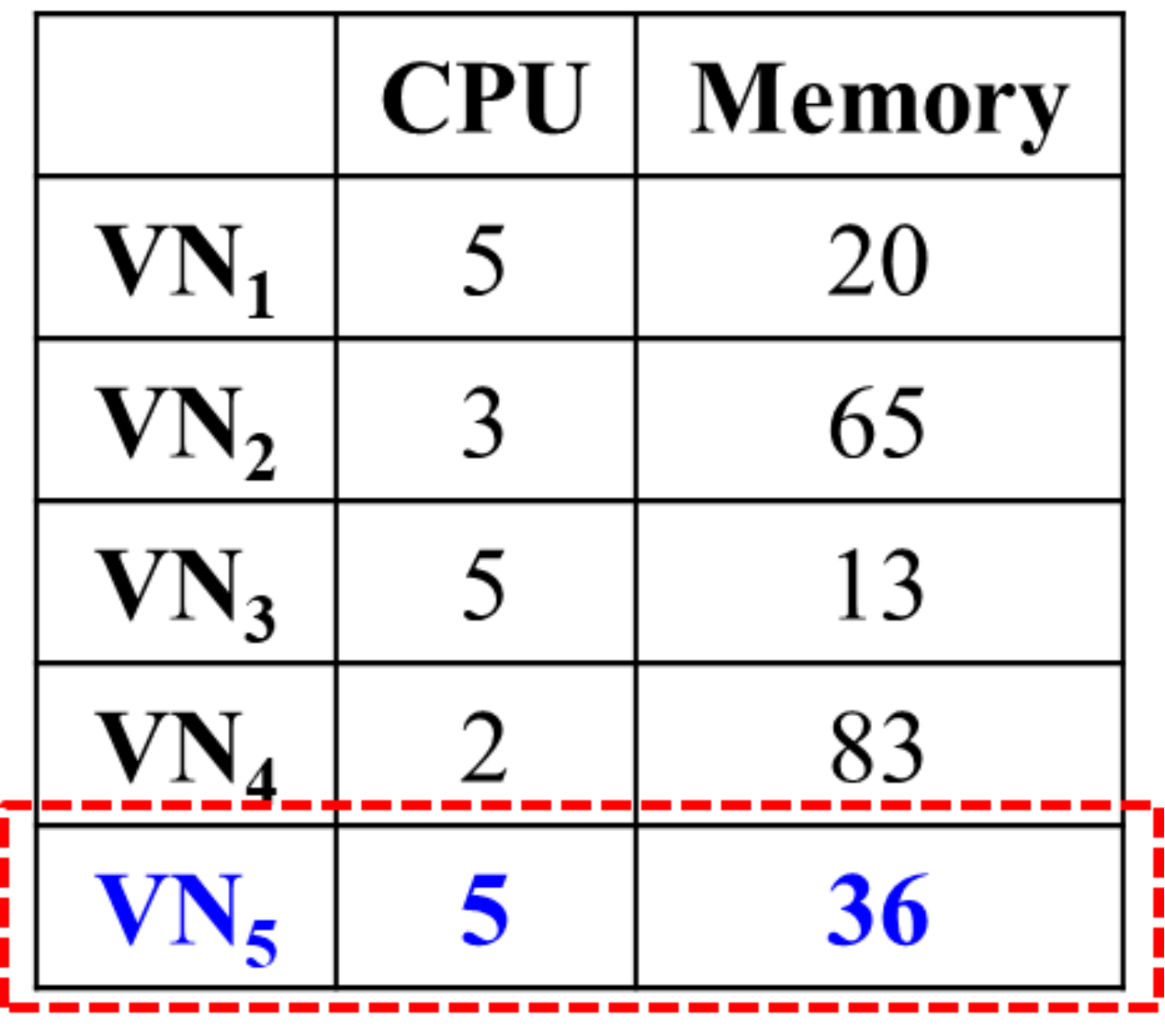}
        \label{fig:Motiv:Util_New_VN_req}
    }
\\
    \subfigure[Two possible solutions, \emph{Uneven distribution} and \emph{Even distribution}.]
    {
        \includegraphics[width=0.48\linewidth]{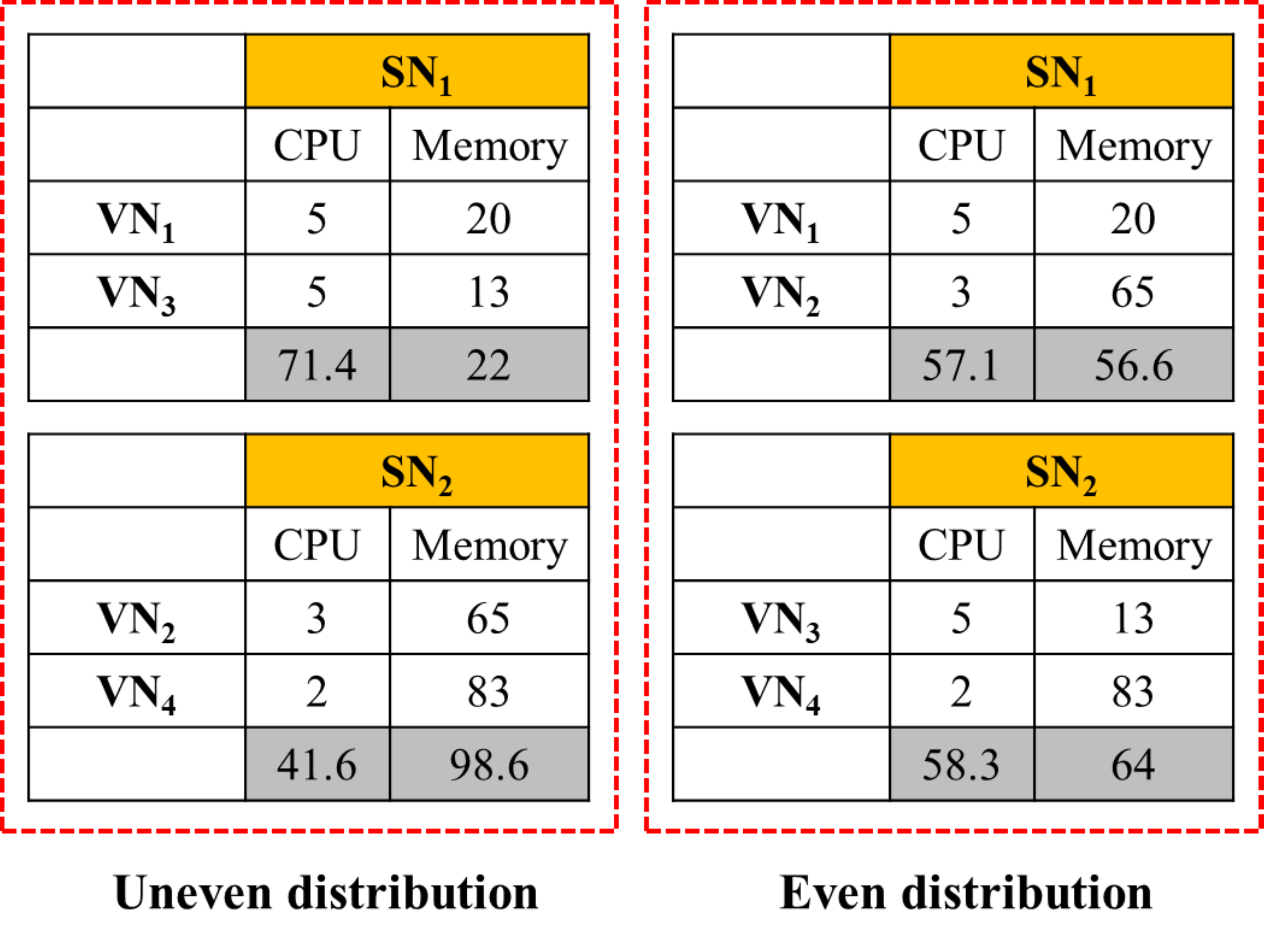}
        \label{fig:Motiv:Util_Sol}
    }~
    \subfigure[Embedding new $VN_5$ in \emph{Uneven distribution} and \emph{Even distribution}.]
    {
        \includegraphics[width=0.48\linewidth]{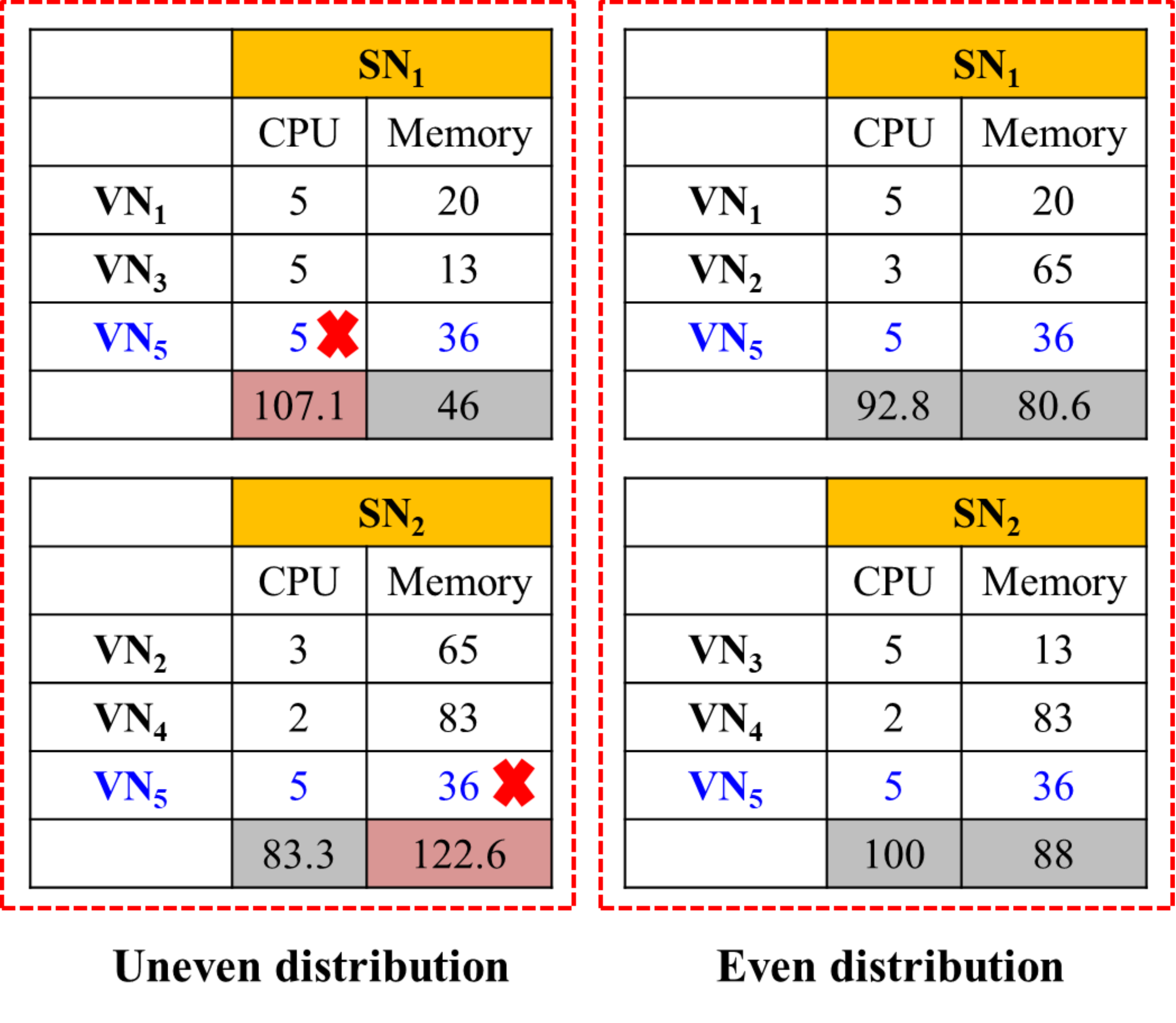}
        \label{fig:Motiv:Util_New_Sol}
    }
    \caption{A motivational example on computing resource utilization.}
    \label{fig:Motiv_comp_resrc_util}
\end{figure}

Furthermore, the workload distribution of substrate resources to
VNs plays an important part in cloud resource management.
Normally, it is preferable to have uniform substrate resource
allocation to VNs across the cloud. Fig.
\ref{fig:Motiv_comp_resrc_util} illustrates the possible problem
scenario that may arise under the uneven workload distribution. Let CSP is comprised of $SN_1$ and $SN_2$ with
maximum CPU capacity of $14$ and $12$ units, respectively and the
maximum memory capacity of $150$ units each, as shown in Fig.
\ref{fig:Motiv:Util_VN_req}. Let CSP receives
VN requests $VN_1, VN_2, VN_3,$ and $VN_4$ with the
CPU demand of $5$, $3$, $5$, and $2$ units, respectively and
memory demand of $20$, $65$, $13$, and $83$ units,
respectively as shown in Fig. \ref{fig:Motiv:Util_VN_req}. The CSP
may embed VN requests in two different ways depicted as
\emph{Uneven distribution} and \emph{Even distribution} as shown
in Fig. \ref{fig:Motiv:Util_Sol}. In case of \emph{Uneven
distribution}, CSP ends up allocating $71.4\%$ and $41.6\%$ units
of the CPU resources, and $22\%$ and $98.6\%$ units of the memory
resources of $SN_1$ and $SN_2$, respectively as shown in Fig.
\ref{fig:Motiv:Util_Sol}. However, in \emph{Even distribution},
the resources are evenly distributed among the SNs. The benefit of
\emph{Even distribution} can be realized as shown in Fig.
\ref{fig:Motiv:Util_New_Sol}. On the arrival of new virtual
network request $VN_5$ with CPU and memory demand of $5$ and $36$
units, respectively as shown in Fig.
\ref{fig:Motiv:Util_New_VN_req}, it becomes infeasible to embed
the $VN_5$ in case of \emph{Uneven distribution}. On the contrary,
$VN_5$ can be successfully embedded onto $SN_1$ or $SN_2$ in case
of \emph{Even distribution} as shown in Fig.
\ref{fig:Motiv:Util_New_Sol}.

It is assumed that at any given time, substrate node can host
multiple VMs of a VN. Under such circumstances, the VN can be
treated as a small graph upon embedding where each vertex would
represent the group of VMs hosted on each SN, and the new virtual
link represents the link between the corresponding SNs. The advantage of such embedding solution is that the network resource demand among VMs in the same substrate node can be ignored as no physical link is involved. However, the computing resources consumed for the communication between the VMs on same SN are not ignored. It is
necessary to distribute the VMs of a VN in such a way that the new
network resource demand of the virtual links can be minimized.
Considering the aforementioned research issues, our goals can be
summarized as to design a reinforcement learning based prediction
models for multi-stage virtual network embedding in cloud data
center to identify and allocate the suitable substrate nodes to VN
requests for subsequent improvement in the overall resource
utilization.

The rest of the paper is organized as follows. Section
\ref{sec:relWorks} describes the related works and the problem is
formulated in Section \ref{sec:sysModel}. The Multi-stage Virtual
Network Embedding (MUVINE) scheme is described in Section \ref{sec:sol} followed
by the knowledge-based AI models in Section \ref{sec:KnowModels}.
The simulation setup and corresponding results are presented in
Section \ref{sec:perfEvaluation} and concluding remarks are made
in Section \ref{sec:conclsn}.

\section{Related Works} \label{sec:relWorks}
In the recent past, a number of VNE schemes have been proposed
considering Data Center (DC) network topologies
\cite{xia2017survey}, resource utilization \cite{2016-5}, energy consumption
\cite{pyoung2018joint}, revenue and profit \cite{Haeri_2017, 2016-VNE}, and survivability of VN \cite{8314665}. A brief summary of the resource allocation schemes is
presented in \cite{zhang2016resource}. The VNE schemes are structured in three broad categories such as Heuristic-based, Nature-inspired learning-based, and Reinforcement learning-based as discussed below.

\subsection{Heuristic-based}

The minimization of a total number of required SNs subsequently improves the overall computing resource utilization. In \cite{sahoo2018lvrm}, a link-based virtual resource management algorithm is proposed considering the substrate network parameters such
as the load, round trip time, and energy consumption. However, it
is observed that resource utilization can be optimized further by
analyzing the historical virtual network requests including the
actual resource usage, substrate resource availability, etc. In \cite{2016-5}, the degree of the nodes and their cluster
coefficient are used to find out the importance of the respective
node and thereby to find out the set of suitable substrate nodes
to embed the VN requests. Although the network topology is taken
into consideration for the VNE, the dependencies among the VNs are
not considered. This results in inefficient VNE in terms of
resource utilization. In \cite{Aral201689}, multiple VM clusters
are mapped in a federated cloud environment. The goal of the
\cite{Aral201689} is to minimize the network latency with
maximization of acceptance ratio. However, authors in \cite{Aral201689}
overlook the fact that geographically distributed cloud
environment incurs the extra cost in terms of time. In
\cite{2016-3}, location preference of the users is considered to
design the VNE algorithm. The VNE problem is formulated as a graph
bisection problem, and VMs and the virtual links are mapped in an
integrated manner. In order to reduce the substrate server failure
impact, VMs are mapped onto separate substrate servers. However,
such mapping results in inefficient network resource utilization.


To provide full network bandwidth guarantee in case of multiple substrate network failure is a challenging issue in cloud management. In \cite{8314665}, a survivable VNE algorithm is proposed by formulating the optimal solution as a quadratic integer program. However, the VNE scheme \cite{8314665} often requires network bandwidth in excess to that of the actual demand. The excess resource requirement can further be minimized by analyzing the actual network resource requirement, which can be predicted by investigating the historical information. In \cite{2016-7-Yin}, an I-VNE algorithm is proposed considering the link interference. The I-VNE algorithm embeds the VNs by considering the temporal and the spatial network topology. The primary advantage of I-VNE algorithm is VNs are embedded with low interference. In \cite{Haeri_2017}, a Multi-commodity flow and shortest-path approaches are followed to embed VNs onto SNs with the objective to maximize the profit of infrastructure provider. The proposed scheme \cite{Haeri_2017} maps the VMs and the virtual links without splitting the substrate path. However, future resource demand and historical information are not considered, which restricts the profit maximization up to certain extent. In \cite{2016-VNE}, a policy-based VNE architecture is proposed that follows network utility maximization approach and separates the high-level goals from the mapping mechanisms of VM and virtual links.

\subsection{Nature-inspired learning-based}
The non-deterministic polynomial hard (NP-hard) characteristic of a VNE problem has encouraged several researchers to explore the nature-inspired learning-based algorithm. In \cite{song2019divide}, a divide-and-conquer evolutionary algorithm (ODEA) is designed for a large-scale VNE. Large VNs are decomposed into overlapping sub-VNs followed by optimization of each sub-VN using the sub-graph mapping procedure. The decomposition of large VNs into sub-VNs makes the ODEA scalable. However, it likely increases the network complexity of the VNE solution, when multiple sub-VNs are embedded on SNs across the network. Besides, the integration of sub-VNs is also a challenging task and it further complicates the VNE solution. 

In \cite{song2019constructive}, a constructive particle swarm optimizer is designed for the VNE (CPSO-VNE). In CPSO-VNE, the VNE process carried out in a step-by-step mapping of each node along with its adjacent virtual links. The CPSO-VNE uses adjacency information to closely embed the virtual nodes in SN topology. However, CPSO-VNE still uses the heuristic information to guide the path information and search. In \cite{song2019distributed}, a distributed VNE system is designed with historical information and meta-heuristic approaches. The particle swarm optimization technique is used to improve the VNE optimization capacity. However, the VNE scheme \cite{song2019distributed} fails to survive the physical failures such as node or link failure. In \cite{gao2013multi}, a multi-objective ant colony system (MOACS) algorithm is proposed for VM placement in the cloud. The MOACS obtains a set of non-dominated solutions that simultaneously minimize total resource wastage and power consumption.
%
%
%
\subsection{Reinforcement learning-based}
More recently, Reinforcement Learning (RL) based VNE is explored. In \cite{2018_reinforcement_VNE}, RL-based VNE algorithm is proposed considering the historical data of the virtual network requests. However, only the resource capacity is considered as a prominent feature for the analysis, which does not ensure state-of-the-art improvement in results. In addition to the resource capacity, features such as VNs arrival time, execution time, resource demand, and the actual resource usage play critical role in deciding the efficient VNE solution. In \cite{2018_adaptive}, the RL-based framework is designed for adaptive resource allocation and provisioning in multi-service cloud environment. The RL-based framework \cite{2018_adaptive} adapts to the system changes such as service cost, resource capacity, and resource demand and it performs well in a multi-service cloud environment. However, it specifically focuses on resource provisioning to satisfy the service level agreement (SLA) of multiple clients. 

Considering the underlying complexity of the cloud environment, it is required to embed intelligence to predict the consequences for the smooth functioning of cloud operations. In \cite{zhang2018intelligent}, an intelligent cloud resource management architecture is designed using deep reinforcement learning. The very purpose of the proposed cloud resource management architecture is to enable cloud to efficiently and automatically negotiate the most suitable configuration derived from the underlying complexity. In \cite{wang2019coordinated}, a coordinated two-stages VNE algorithm is designed using the RL. A pointer network is used to find a node mapping policy and shortest path algorithm is used to embed the links. The VNE algorithm of \cite{wang2019coordinated} mainly focuses on the CPU and bandwidth as resources to optimize. However, it is challenging to design an efficient VNE algorithm without considering the memory as a resource. In \cite{yao2018rdam}, historical network request data and policy-based RL is designed for node mapping. The overall objective of policy-based RL \cite{yao2018rdam} is to maximize the revenue by embedding nodes and links using convolutional neural network. Similarly, in \cite{zhang2018efficient}, a deep learning model is designed to predict the cloud workload for the industry informatics. The workflow schedule is an integral operation of any cloud computing environment. The primary objective of workflow scheduling is to optimize the processing cost and thereby enable the cloud to efficiently provide the services to the end-user and at the same time remain profitable. In \cite{sagnika2018workflow}, a Bat Algorithm is employed to efficiently schedule the data-intensive workflow applications. Similarly, the self-management of cloud resources is an important yet challenging objective and needs further investigation. In \cite{mijumbi2015neuro}, a dynamic, decentralized, and coordinated neuro-fuzzy approach is proposed to self-manage the substrate network resources.

\section{Problem Formulation} \label{sec:sysModel}
In this section, the mathematical formulation of the substrate
network and virtual network requests is presented. The brief
summary of notations used in the paper is presented in Table
\ref{table:notation}.

\subsection{Substrate network}

The typical cloud environment is comprised of a large number of SNs often referred to as the physical machines. It is
assumed the SNs are connected through a switch-centric network
topology. The resources are divided into two categories such as
computing and network. The computing resource refers to CPU and
memory. The network of SNs is denoted by $G^s(N^s,
E^s)$, where $N^s = \{S_1, S_2, S_3, ..., S_m\}$ be the set of $m$
SNs, and $E^s$ be the set of the substrate links. Each
SN is denoted as $S_i \in N^s, 1 \le i \le m$. The maximum
computing resource capacity of type $x \in \{CPU, memory\}$
available at each $S_i$ is denoted by $C_i^x$; whereas the
remaining computing resource availability is denoted as $A_i^x$.
For example, $A_5^{CPU}$ indicates the remaining CPU resource
availability of the SN $S_5 \in N^s$. Similarly, $L_{ij}^a$ be the network bandwidth availability of substrate link between $S_i$ and $S_j$. Using the $L_{ij}^a$, the network bandwidth
availability of the SN $S_i$ denoted as $L_i^a$ can be calculated
as follows.

\begin{equation}
L_i^a = \max_{\forall S_j \in N^s}\{L_{ij}^a\}, \quad i\ne j
\end{equation}

It is assumed that the SNs may be equipped with different amount of resources. Hence, the $C_i^x$ and $A_j^x$ may differ for any substrate nodes $S_i$ and $S_j$, $i\ne j$. Let $L_{ij}^l$ be the total number of physical links in the shortest path between $S_i$ and $S_j$. A shortest path algorithm, such as Dijkstra's algorithm, can be used to find the shortest path with minimum number of intermediate physical links. 

\begin{table}
    \begin{center}\caption{List of notations}
        \begin{tabular}{|c|l|}
            \hline
            \textbf{Notation}   & \textbf{Description}                                                           \\ \hline
            $G^s(N^s, E^s)$    & Substrate network graph                                                         \\ \hline
            $N^s$		&Set of SNs
            \\ \hline
            $E^s$		& Set of substrate links
            \\ \hline
            $G^v(N^v, E^v)$ & Virtual network graph                                              \\ \hline
            $N^v$		& Set of VMs
            \\ \hline
            $E^v$		& Set of virtual links
            \\ \hline
            $n$         & Total number of VMs in the current VN                                                    \\ \hline
            $m$          & Total number of SNs                                                      \\ \hline
            $S_i$         & Represents an SN, $1 \le i \le m$                                                   \\ \hline
            $V_j$        & Represents a VM, $1 \le j \le n$                                                    \\ \hline
            $C_i^x$       &Maximum available resource of type $x$ at $S_i$, $x \in \{CPU, memory\}$ \\ \hline
            $A_i^x$       & Amount of resource of type $x$ available at $S_i$, $x \in \{CPU, memory\}$  \\ \hline
            $D_j^x$       & Resource demand of type $x$ by $V_j$, $x \in \{CPU, memory\}$              \\ \hline
            $L_{ij}^a$     & Amount of network resource available between $S_i$ and $S_j$ \\ \hline
            $L_{ij}^d$     & Network resource demand  by the virtual link between $V_i$ and $V_j$ \\ \hline
            $L_{ij}^l$		& The total number of physical links in the shortest path between SN $S_i$ and $S_j$      \\ \hline
            $L_i^a$       & Amount of network resource available at $S_i$                   \\ \hline
            $L_j^d$       & Network resource demand  by $V_j$                   \\ \hline
            $\alpha_j$  & Class of $V_j$ \\ \hline
            $\alpha$  & Class of VN request \\ \hline
            $\beta$  & Priority value of VN request \\ \hline
            $\kappa_i^j$  & Boolean value to indicate $V_j$ assigned to $S_i$
            \\ \hline
            $\omega_n$		& Resource preference constant
            \\ \hline
            $\phi_j$		& Start time of $V_j$
            \\ \hline
            $\chi_j$		& End time of $V_j$
            \\ \hline
            $rC_i^x$		& Actual resource consumption of type $x$ by $V_j$
            \\ \hline
            $\Phi$		& Aggregate CPU demand of $G^v(N^v, E^v)$
            \\ \hline
            $\Psi$		& Aggregate memory demand of $G^v(N^v, E^v)$
            \\ \hline
            $\phi_v$		& Start time of $G^v(N^v, E^v)$
            \\ \hline
            $\chi_v$		& End time of $G^v(N^v, E^v)$
            \\ \hline
            $rC_v^x$		& Actual resource consumption of type $x$ by $G^v(N^v, E^v)$
            \\ \hline
            $\ell$			& Binary label of VN training dataset  \\ \hline
            $\zeta$			& Number of features in VN dataset		\\ \hline
            $Q(s_t, a_t)$	& State-action value function		\\ \hline
            $\mathbb{R}(G^{v}(N^{v},E^{v}))$	& Reward of $G^{v}(N^{v},E^{v})$  \\ \hline
            $\mathbb{H}$	& Decision variable  \\ \hline
        \end{tabular}
        \label{table:notation}
    \end{center}
\end{table}

\subsection{Virtual network}
It is assumed that the cloud users send the request in the form of a virtual network (VN) represented as an undirected weighted graph
denoted as $G^v(N^v, E^v)$. Each $G^v(N^v, E^v)$ is comprised of a set of $n$
number of VMs denoted as $N^v = \{V_1, V_2, V_3, ...., V_n\}$ and
a set of virtual links $E^v$. Each VM, denoted by $V_j, 1 \le j
\le n$, is associated with computing resource demand parameter
denoted by $D^x_j$. Here, $x \in \{CPU, memory\}$ represents the
computing resource type. Similar to the VM resource demand, each
virtual link between VM $V_i$ and $V_j$ is associated with network
bandwidth demand denoted by $L_{ij}^d$. Based on the value of
$L_{ij}^d$, the network bandwidth demand of a particular VM $V_i$
can be calculated as follows:

\begin{equation}
L_i^d = \sum_{\forall V_j \in N^v}{L_{ij}^d}
\end{equation}

Let $L_{ij}^l(L_{u_1u_2}^d)$ be the number of physical links in the shortest path between SN $S_i$ and $S_j$ provided that this shortest path fulfills the bandwidth demand of virtual link between virtual node $V_{u_1}$ and $V_{u_2}$. The VNs are assumed to be static in nature. This infers that the
resource demand and the structure of the VNs do not change upon
the embedding. The end time of the virtual network is unknown.
Hence CSP has no information regarding the time duration the
computing and network resources need to be reserved for each VN.
The boolean variable $\kappa_i^j$ is used to indicate if $V_j$ is assigned to substrate node $S_i$. Mathematically,

\begin{equation}
\kappa_i^j = \left\lbrace
\begin{array}{c l}
1 & \text{if } V_j ~\text{is assigned to substrate node } S_i \\
0 & \text{Otherwise}
\end{array}\right.
\end{equation}

\subsection{VN classes and priority}
The incoming requests are classified into three different classes
based on the delay sensitiveness: Class 1, Class 2, and Class 3.
The class $\alpha_j, 1 \le j \le n$ of any $V_j$ is decided by
the user, whereas the class $\alpha$ of any $G^v(N^v, E^v)$ is
decided by the CSP based on the classes of each $V_j \in
G^v(N^v, E^v)$. VMs in Class 1, Class 2, and Class 3 are said to
be highly, moderately, and less delay sensitive, respectively. CSP
strictly follows the resource demand of VMs in Class 1; whereas it
is not mandatory for the CSP to strictly fulfill the resource
demand of the VMs in Class 3.

Based on the value of $\alpha$, the priority of the VN is derived, which is denoted by $\beta$. Mathematically,

\begin{equation}
\beta = \left\lbrace
    \begin{array}{c l}
    1 & \text{if }\alpha = \emph{Class 3}, \text{ less delay sensitive VN} \\
    2 & \text{if } \alpha = \emph{Class 2}, \text{ mderately delay sensitive VN} \\
    3 & \text{if } \alpha = \emph{Class 1}, \text{ highly delay sensitive VN} \\
    \end{array} \right.
\end{equation}
During the substrate resource allocation, VNs with higher priority value such as $\beta=3$, are given higher
preference. On the
contrary, VNs with less priority value are given less importance
in the resource allocation process.

\subsection{Objective function}
Considering the above scenario, the main objective of our
proposed MUVINE scheme is to maximize the
mean percentage of the resource utilization based on the
predictions through the reinforcement learning. The resource maximization must be done along with the
maximization of the acceptance rate. Here, the acceptance rate
refers to as the percentage of the VNs that are accepted without
violating the SLA and resource demands out of the total requested
VNs. Mathematically, the objective function can be written as
shown in Eq. \ref{eq:objective}.


\begin{align}
\label{eq:objective}
maximize~Z = & \left[\sum_{\forall x \in \{CPU, memory\}}\sum_{\forall S_i \in N^s} \sum_{\forall V_j \in N^v} \frac{\kappa_i^j}{m} * \right.\nonumber \\  
& \left. \qquad \left(\omega_x*{\frac{C_i^{x} - A_i^{x} + D_j^{x}}{C_i^{x}}}\right)\right] \nonumber \\
& + \left[ \sum_{\forall S_{i_1}, S_{i_2} \in N^s} \sum_{\forall V_{u_1}, V_{u_2} \in N^v} \frac{\omega_{n}* \kappa_{i_1}^{u_1} * \kappa_{i_2}^{u_2}}{L_{i_{1}i_{2}}^l(L^d_{u_{1}u_{2}})}\right]
\end{align}

\textbf{Constraints:}

\begin{equation}\label{eq:const:one2one}
\sum_{i=1}^{m} \kappa_i^j = 1, \quad \forall V_j \in N^v
\end{equation}


\begin{equation}\label{eq:const:feasibleResrcDmnd}
\forall S_i \in N^s, \nexists V_j \in N^v, D_j^x \ge A_i^x, L_j^d \ge L_i^a, x \in \{CPU, memory\}
\end{equation}


\begin{equation} \label{eq:const:dmnd_avail}
\text{if}~\kappa_i^j = 1, \quad 0 < D_j^x < A_i^x \text{ and } 0 < L_j^d < L_i^a
\end{equation}

\begin{equation}\label{const:omega}
\omega_n + \sum_{x\in \{CPU, memory\}}\omega_x = 1
\end{equation}


The objective function shown in Eq. \ref{eq:objective} is of two folds, where the first part primarily focuses on maximization of server resources (CPU, memory) utilization and the second part focuses on minimizing the number of physical links used. To minimize the number of physical links, it is encouraged to embed the VMs on the nearby SNs. For this, the Dijkstra algorithm is used to find the shortest path between two SNs. By minimizing the number of physical links, a small amount of network resource is saved for each VN request, which helps increasing the acceptance rate. The obtained solution of the above objective function must satisfy the constraints as shown in Eq. \ref{eq:const:one2one} - \ref{const:omega}. The constraints are described as follows.
\begin{itemize}
	\item  The constraint shown in Eq. \ref{eq:const:one2one} ensures each VM is embedded exactly onto one SN. This implies that no splitting will be done with the VM. Further, it does not guarantee how many number of VMs from one VN request will be embedded onto a single SN.
	\item The constraint shown in Eq. \ref{eq:const:feasibleResrcDmnd} ensures that VN request is not comprised of any VM with the resource demand greater than the resource availability on any of the SN. To be specific, there exists at least one SN for each VM that fulfills the resource demand. 
	\item The constraint shown in Eq. \ref{eq:const:dmnd_avail} refers to embedding the VM onto the SN where the resource demand is less than the resource availability, which also indicates that a positive resource demand value is associated with all VMs and virtual links.	
	\item $\omega_x$ and $\omega_n$ are the constants introduced to assign a preference value to different types of resources such as CPU, memory, and network bandwidth. Constraint as shown in Eq. \ref{const:omega} ensures that the sum of all the constants is $1$.
\end{itemize}

\section{The Multi-stage Virtual Network Embedding (MUVINE) scheme}\label{sec:sol}
In this section, the MUVINE scheme is described, which is designed on the top of the reinforcement learning based predictions. It is assumed that the
CSP (agent) learns by exploring the existing resources of the
cloud data centers (environment). The purpose of exploration is to
learn series of actions from the cloud environmental and ascertain an optimum action that maximizes the total cumulative reward of the CSP. The typical cloud environment dataset is described in detail as follow.

\subsection{Cloud environment dataset}\label{his:data}
Let $m$ be the number
of SNs with heterogeneous computation and network capacity in a cloud data center. Let multiple applications arrive
at the CSP in the form of VNs, which in turn be represented as the set of
interconnected VMs. It is to be noted that each VN may have
different graph structure and two VNs with the same graph
structure may have VMs with different resource requirement. The historical information is comprised of the set of
features and labels. The features represent the parameters of VNs,
VMs, and SNs.
\subsubsection{\textbf{VM feature set}} \label{VM_feature_set}
Let $G^v(N^v, E^v)$ is comprised of $n$ VMs. Without losing the generality, Let $V_j \in G^v(N^v, E^v)$ contains set of user-specific computational capacity parameters such as CPU and memory demand of the $V_j$ represented as $D_j^{CPU}$ and $D_j^{memory}$, respectively. Besides, each $V_j \in G^v(N^v, E^v)$ also contains the user-specific parameters such as class of the $V_j$, priority of $V_j$ represented as $\alpha_j$ and $\beta_j$, respectively. In addition to the user-specific parameters, CSP also records parameters such as start time of $V_j$ represented as $\phi_j$, end time of $V_j$ represented as $\chi_j$, actual resource consumption of type $x$ by $V_j$ represented as $rC_j^{x}$, where $x \in \{CPU, Memory\}$.
\subsubsection{\textbf{VN feature set}} \label{VN_feature_set}
Similar to $V_j$, the $G^v(N^v, E^v)$ is also associated with user-specific parameters and CSP recorded parameters. The user-specific parameters associated with $G^v(N^v, E^v)$ are aggregate CPU demand represented as $\Phi$, aggregate memory demand represented as $\Psi$, class of $G^v(N^v, E^v)$ represented as $\alpha$, priority of $G^v(N^v, E^v)$ represented as $\beta$. The CSP recorded parameters associated with VN $G^v(N^v, E^v)$ are start time of $G^v(N^v, E^v)$ represented as $\phi_v$, end time of $G^v(N^v, E^v)$ represented as $\chi_v$, actual resource consumption of type $x$ by  VN $G^v(N^v, E^v)$ represented as $rC_v^{x}$, where $x \in \{CPU, Memory\}$.
\subsubsection{\textbf{SN feature set}}
At any given time in the cloud, the SNs represented as $S_i$, $1 \le i \le m$ may contain available resource of type $x$ represented as $A_i^{x}$, where $x \in \{CPU, memory, network\}$. Moreover, the CSP also records and maintain the available clock rate represented as $A_i^{Cr}$.

The aforementioned set of parameters are referred to as the aggregate set of features recorded by CSP for each VN execution and together constitute it as a historical dataset.

\subsection{Multi-stage prediction architecture}
\begin{figure}
    \centering
    \includegraphics[width=0.7\linewidth]{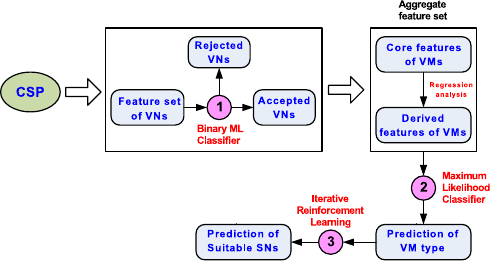}
    \caption{Multi-stage prediction architecture}
    \label{fig:multi-stagearchitecture}
\end{figure}

The multi-stage prediction architecture for VNE is presented in
Fig. \ref{fig:multi-stagearchitecture}. A binary ML classifier
followed by Maximum Likelihood Classifier (MLC) and iterative
reinforcement learning are employed to select the set of most
appropriate SNs to embed the VN requests. The prediction
architecture can be described as follows.

In the first stage, the acceptability of the VN request into the
cloud data center is evaluated. A binary ML classifier is designed
to classify VN requests into "accepted" or "rejected". The binary
classifier is built and trained on the top of the historical
dataset of VNs comprised of features as described in Section
\ref{VN_feature_set} and an observed binary value of label $\ell
\in \lbrace accepted, rejected \rbrace$. It is to note that the
label $\ell$ is CSP assigned and it is based on the actual
outcomes observed for the completed VN requests. Using the
aforementioned binary classifier, each real-time VN request is
classified. For each "accepted" VNs, the system enters into the
second stage of the prediction.

For each accepted VN, the type of each member VM is predicted. In
particular, for a given $G^v(N^v, E^v)$, each $V_j \in
G^v(N^v, E^v)$, $1 \leq V_j \leq n$ is classified into three
categories defined as "CPU intensive", "GPU intensive", and
"memory intensive" based on the VM configuration and its resource
analysis. From the historical VM dataset comprised of features
described in \ref{VM_feature_set}, a multi-class (three)
classifier is trained to estimate the VM type. It is to note that
VM requests often arrive with partial information such as CPU and
Memory demand represented as $D_j^{x}$, where $x \in \{CPU,
Memory\}$, VM class $\alpha_j$, VM priority $\beta_j$. The
aforementioned parameters are together called the core VM
features. However, certain information may not be made available at
the time of VN request arrival and need to be estimated such as end time of VM $\chi_j$,
actual CPU consumption, actual memory consumption together denoted
as $rC_{j}^{x}$, where $x \in \{CPU, memory\}$ etc. Such features are referred to as the derived
features. The regression analysis is performed to estimate the
most appropriate value of derived features based on the available
core features. The core features and derived features together constitute the "aggregate
feature set".

Upon the classification of each $V_j \in G^v(N^v, E^v)$, $1 \leq
V_j \leq n$ into the predefined VM type, an iterative
reinforcement learning is applied for the embedding. Using the Substrate network environment, which represents the current state of the cloud data center,
iterative reinforcement learning predicts the most appropriate
SNs with sufficient resource availability to embed the VMs in a manner
that balances the resource utilization across the cloud data center. Usually, a huge number of parameters are required
to optimize on a real-time basis to efficiently manage the cloud data center resources. Since the cloud data center resource configuration dynamically changes with each VN request embedding, it becomes challenging to predict the resource availability on real-time basis for the embedding of incoming new VN requests. Among the
RL approaches, Q-learning and SARSA
(State-Action-Reward-State-Action) are highly relevant to address
the VNE problem. The Q-learning being an off-policy method follows
the greedy approach and takes the action, i.e., the selection of
the suitable SNs to embed the VMs based on the maximum reward that
it can achieve in the current state. However, a greedy approach can
lead to the local optimum VNE rather than the global optimum one.
Hence, an on-policy RL method SARSA is adopted. Throughout the
paper, the actions are referred to as the VNE in the context of
the RL. In SARSA, the policy is continuously updated based on the
actions it takes. Contrary to the Q-learning, SARSA effectively
balances between the exploration and exploitation of the given
environment to learn the optimum policy decision.


\subsection{Features and predictors}\label{sec:2ndStage}
The features and predictors are integral parts of any ML-based
algorithm. In this paper, we define two types of features called
as core and derived features with respect to VNE in a cloud data
center. Often it is considered that the features available at the
time of training an ML algorithm, the same set of features are
assumed available at the time of testing. However, in many
practical applications, it is not the case and the value of a few
features are not made available until the execution of the process
is concluded. In the context of the VNE, the parameters such as
end time of VN and VM, actual resource utilization of VN and VM,
are few of the parameters that are not available beforehand and
therefore supervised ML algorithms are difficult to apply. In
order to tackle the aforementioned problem of feature
availability, training is performed in two sequential steps. In
the first step, the core features are used and an estimated value
of the rest of the derived features are predicted for a given VN
request. Later, using the combined set of core and derived
features also known as the aggregate set of the features, an
ML classifier is trained to obtain the results.
The definition of core and derived features is described as
follow.

\begin{itemize}
    \item \textbf{Core feature set}: The set of primary features available at the time of VN and VM request.

    \item \textbf{Derived feature set}: The set of features, which are not available at the time of VN and VM request, but are derived from the set of core features.
\end{itemize}

The predictors are commonly known as the set of labels that the
algorithm predicts for the given test sample. For the VN feature
set, the predictor is a binary value representing the notion
"accepted" and "rejected". For the aggregate feature set of VM,
the predictor is the multi-class label representing the class "CPU
intensive", "GPU intensive", and "memory intensive".

\section{Knowledge-based model designing} \label{sec:KnowModels}
In this section, we design knowledge-based models for efficient VNE.
\subsection{\textbf{Binary VN classifier}} \label{sub:SVMClassifier}
In this section, we design a supervised support vector binary
classifier to classify the VN requests into "accepted" or
"rejected" using the historical VN dataset. The support vector machine (SVM) is a simple yet robust supervised non-probabilistic binary classifier. The objective of SVM is to find a unique maximum margin hyperplane in a multi-dimensional space that maximizees the margin between two classes and efficiently distinguishes VN request into "accepted" and "rejected". Given the labelled training dataset of VN requests, the SVM generates an optimal hyperplane that categorize the new VN request. On the top of the
historical VN dataset, the proposed support vector binary
classifier is trained and a representative $\zeta -1$ dimensional
hyperplane is constructed in such a way that with respect to the
support vectors in class "accepted" and "rejected", the
maximum-margin hyperplane is derived. Here, $\zeta$ represents the
number of features in VN dataset. Let us assume that the $\varpi$
represents the VN dataset comprised of linearly separable labeled
training samples. For simplicity, we redefine VN $G^v(N^v, E^v)$
as $VN_v$.The $\varpi$ can be represented as follow, $\varpi =
\{(VN_1,l_1), (VN_2,l_2),...,(VN_k,l_k)\}$, where $k$ represents
the total number of samples in $\varpi$ and $k > 0$. It is to note
that $VN_i \in \varpi$, $i=\{1,2,...,k\}$ represents $i^{th}$
training VN request comprised of $\zeta$ dimensional feature
vector $VN_{f}$. The $l_i \in \varpi$, $i=\{1,2,...,k\}$
represents the output class label of $i^{th}$ training sample
assigned by CSP. Let us say $l_{i}=1$ represents the output class
label "accepted" and $l_{i}=-1$ represents the output class label
"rejected". The ultimate goal of the proposed binary classifier is
to derive an optimum hyperplane that partitions the training data
set $\varpi$ into subsets "accepted" and "rejected". The subset
"accepted" can be defined as $\varpi^{+1} = \lbrace VN_i \in
\varpi | l_i =+1 \rbrace$, and the subset "rejected" can be
represented as $\varpi^{-1} = \lbrace VN_i \in \varpi | l_i =-1
\rbrace$, where $\varpi^{+1} \cap \varpi^{-1} = null$. The
$l_i=+1$ and $l_i=-1$ represent the class "accepted" and
"rejected", respectively.

For simplicity, let us represent $\aleph _{w,c}(\varpi)$ as a linear classifier. The $\aleph_{w,c}(\varpi)$ can be defined as shown in Eq. \ref{SVMclassifier}.

\begin{align}
\label{SVMclassifier}
\aleph _{w,c}(\varpi)=w^{T} \times VN_f+c,~~~ \forall VN_i \in \varpi
\end{align}

Here, $\aleph _{w,c}(\varpi)$ represents a classification hyperplane,
$VN_{f}$ represents $\zeta$-dimensional feature-rule vector corresponding
to $VN_{i}$, $w$ represents the $\zeta$-dimensional weight vector
associated with feature-rule vector $VN_{f}$, and $c$ represents
the bias. The $\zeta$-dimensional weight vector $w$ is a learning
vector, which is learned by the classifier from the
training dataset $\varpi$.

Three hyper-planes are defined for the classification purpose. The
optimized classification hyperplane is represented as $\aleph
_{w,c}(\varpi)=w^{T} \times VN_f+c=0$. The hyperplane for the
nearest data point also known as support vector in class
"accepted" ($l_i=+1$) can be represented as $\aleph
_{w,c}(\varpi)=w^{T} \times VN_f+c=+1$, and the hyperplane for the
nearest support vector in class "rejected" ($l_i=-1$) can be
represented as $\aleph _{w,c}(\varpi)=w^{T} \times VN_f+c=-1$. The
distance between hyperplane $\aleph _{w,c}(\varpi)=+1$ and $\aleph
_{w,c}(\varpi)=-1$ can be defined as $\dfrac{2}{||w||}$ and is
maximized under the constraint $l_i(w^{T \times }VN_f+c)\ge1$, for
each $VN_i \in \varpi$.

In many instances, it is likely that the historical training dataset may not be linearly separable. To tackle such situations, the SVM is
extended to include the hinge loss function
$max(0,1-l_i(w^{T} \times VN_f+c))$. In this case, the optimization
problem can be defined as shown in Eq. \ref{softmargin}, where
$\lambda$ decides the margin size and at the same time $\lambda$
ensures that the sample $VN_i \in \varpi$ lies in the correct class.

\begin{align}
\label{softmargin}
min\bigg(\bigg[\dfrac{1}{m}\sum_{i=1}^{m}&max(0,1-l_i(w^{T} \times VN_f+c))\bigg]+\lambda||w||^2\bigg),&&\\\nonumber & where~m=number~of~samples
\end{align}

Upon the completion of support vector classifier training, the derived
hyperplane is employed to classify the unseen VN test set $\varpi_{Te}$. The VN request $VN_{i} \in \varpi_{Te}$ that classifies into
class "accepted" (+1) is considered as eligible.

\subsection{Radial Basis Regressors (RBR)} \label{RBF}
The "accepted" VNs are further analyzed to build the aggregate feature set of VMs comprised of core features and derived features. 
Let $\mathbb{D}$ be the historical VM dataset of $N$ data points represented as $(x_n$,$y_{n1}$,$y_{n2}$,...$y_{nk})$, $n=\lbrace 1,2,..,N\rbrace$. Each data point $(x_n,y_{n1},y_{n2},...y_{nk}) \in \mathbb{D}$ represents a unique historical sample of training dataset. Here, $x_n$ represents the core feature set of $n^{th}$ sample, and $y_{n1},y_{n2},...y_{nk}$ represents the observed output value of $k$ derived features in $n^{th}$ training sample. For the estimation of $y_{n1},y_{n2},...y_{nk}$, the regression approach is adopted. Instead of estimating all of the $k$ derived features in a single regression model, $k$ individual regression models are designed to estimate each individual derived feature. Considering the non-linearity nature of the training dataset $\mathbb{D}$, instead of employing linear regression, radial basis function is applied.

The radial basis regressors (RBR) model can be described as follow. Let $\Theta(x)$ be the RBR model trained on the dataset $\mathbb{D}$ and defined as follow.

\begin{align}
\label{RBF_Regressors}
\Theta(x) = \sum_{n=1}^{N} w_n \times exp\big(-\gamma \lVert x - x_n \rVert^2\big)
\end{align}

Here, $\Theta(x)$ is considered influenced by each training data sample $(x_n,y_{n1},y_{n2},...y_{nk})$, $n=\lbrace 1,2,..\rbrace$ based on the euclidean distance formally defined as $\lVert x - x_n \rVert$. In order to smoothen out the distance parameter $\lVert x - x_n \rVert$, the $\Theta(x)$ model is embedded with the Gaussian basis function $exp\big(-\gamma \lVert x - x_n \rVert^2\big)$. The $\gamma$ is an influential factor. The RBR problem can be transformed into a learning problem to estimate the value of $w_n$, $\forall n=\lbrace 1,2,...,N\rbrace$ in such a way that for each test sample $(x_m,y_{mi})$, the $\Theta(x) = y_{mi}$. The learning problem can be formally defined as follow.

The learning problem is reduced to find the $w_1, w_2,...,w_N$ for Eq. \ref{RBF_Regressors} based on the $\mathbb{D}$ in such a way that the Error is minimized (i.e., $\mathbb{E}=0$). In other words, the objective of the learning problem is to obtain $\Theta(x_n)=y_{mi}$, $\forall n = \lbrace 1, 2,...,N \rbrace$ as defined in Eq. \ref{objective_function}.

\begin{align}
\label{objective_function}
\sum_{m=1}^{N} w_m \times exp\big(-\gamma \lVert x_n - x_m \rVert^2\big)=y_{mi}
\end{align}

The Eq. \ref{objective_function} can be expanded as shown in Eq. \ref{Interpolation}.

\begin{gather}
\label{Interpolation}
\underbrace{\begin{bmatrix}
exp\big(-\gamma \lVert x_1 - x_1 \rVert^2\big) && ... && exp\big(-\gamma \lVert x_1 - x_N \rVert^2\big) \\
exp\big(-\gamma \lVert x_2 - x_1 \rVert^2\big) && ... && exp\big(-\gamma \lVert x_2 - x_N \rVert^2\big) \\
... &&  && ...\\
exp\big(-\gamma \lVert x_N - x_1 \rVert^2\big) && ... && exp\big(-\gamma \lVert x_N - x_N \rVert^2\big)
\end{bmatrix}}_{\mathbf{\Phi}}
\underbrace{\begin{bmatrix}
w_1 \\ w_2 \\ ... \\ w_N
\end{bmatrix}}_{\mathbf{w}}
=
\underbrace{\begin{bmatrix}
y_{m1} \\ y_{m2} \\ ...\\ y_{mN}
\end{bmatrix}}_{\mathbf{y}}
\end{gather}

The exact interpolation or solution may exist if $\Phi$ is invertible. In other words, if $\Phi$ is invertible, the $w$ can be determined as $w = \Phi^{-1} \times y$. By solving the Eq. 13, the estimated value of derived features are obtained.

\subsection{Maximum Likelihood Classifier (MLC)}
Upon obtaining the "aggregate feature set" of incoming VMs, the same are classified into one of the three categories "CPU intensive", "GPU intensive", and "Memory-intensive". Let $\varsigma$ be the aggregate feature set of VM. It is to note that the historical VM training dataset $\vartheta_{tr}$ is comprised of "aggregate feature set" along with output label describing the category of the VMs. The supervised Maximum Likelihood Classifier (MLC) is designed based on the features. Let $cL_j^i$ be the class label of $V_j \in G^v(N^v,E^v)$. Here, $cL_j^{i=1}$, $cL_j^{i=2}$, and $cL_j^{i=3}$ represents the $V_j$ of type "CPU intensive", "GPU intensive", and "Memory-intensive", respectively. The historical VM dataset is considered as a training dataset to learn the feature value for each VM category.

For each incoming $V_j \in G^v(N^v,E^v)$, the corresponding aggregate features $\varsigma$ are obtained using RBR model described in Section \ref{RBF}. Using $\varsigma$, the likelihood probability of class $cL_j^{i=1}$, $cL_j^{i=2}$, and $cL_j^{i=3}$ is calculated using conditional probability. The conditional probability of $cL_j^{i=1}$, $cL_j^{i=2}$, and $cL_j^{i=3}$ can be represented as $p(cL_{j}^{i=1}|\varsigma)$, $p(cL_{j}^{i=1}|\varsigma)$, and $p(cL_{j}^{i=1}|\varsigma)$. For a given aggregate features $\varsigma$ of an incoming VM $V_j \in G^v(N^v,E^v)$, the conditional probability of  class $cL_j^{i=1}$, $cL_j^{i=2}$, and $cL_j^{i=3}$ can be calculated as shown in Eq. \ref{conditional_probability}.

\begin{align}
\label{conditional_probability}
p(cL_{j}^{i}|\varsigma)=p(cL_{j}^{i} | \varsigma^{1}, \varsigma^{2},...,\varsigma^{p}),~i \in \lbrace1,2,3\rbrace
\end{align}
Here, $p$ represents the number of features. The feature-rules $\varsigma^{1}$, $\varsigma^{2}$,...,$\varsigma^{p}$ are assumed conditionally independent to each other. In the presence of conditional independence, Eq.
\ref{conditional_probability} can be rewritten as shown in Eq.
\ref{conditional_independence}.

\begin{flalign}
\label{conditional_independence}
p(cL_{j}^{i} | \varsigma^{1}, \varsigma^{2},...,\varsigma^{p}) \propto p(cL_{j}^{i}, \varsigma^{1}, \varsigma^{2},...,\varsigma^{p})
\end{flalign}

Further, Eq. \ref{conditional_independence} can be expanded as shown in Eq. \ref{conditional_expansion}.

\begin{flalign}
\label{conditional_expansion}
p(cL_{j}^{i} | \varsigma^{1}, \varsigma^{2},...,\varsigma^{p}) \propto p(cL_{j}^{i}) \times p(\varsigma^{1}|cL_{j}^{i}) \times p(\varsigma^{2}|cL_{j}^{i}) \times ... \times p(\varsigma^{p}|cL_{j}^{i})
\end{flalign}

Here, $p(\varsigma^{k}|cL_{j}^{i})$, $k=\{1,2,...,p\}$ represents the probability of obtaining
feature $\varsigma^{k}$ for a given class $cL_{j}^{i}$. The $p(\varsigma^{k}|cL_{j}^{i})$ can be calculated as shown in Eq.
\ref{calculation}.

\begin{flalign}
\label{calculation}
\nonumber p(\varsigma^{k}|cL_{j}^{i}) = \dfrac{1}{\sqrt{2\pi\sigma_{k}^2}} \exp\bigg(\dfrac{-(\varsigma^{k}-\mu_{k})^2}{2\sigma_{k}^2}\bigg)\\
\end{flalign}

The $\mu_{k}$ represents the mean value of feature $\varsigma^{k}$ for class $cL_{j}^{i}$. Similarly,
$\sigma_{k}^2$ represents the standard variance of feature $\varsigma^{k}$ for class $cL_{j}^{i}$. The mean $\mu_{k}$ and standard
variance $\sigma_{k}^2$ are the learning parameters and they can
be estimated from the training set samples of class $cL_{j}^{i}$.

Once the conditional probability model is designed, a Maximum Likelihood Classifier (MLC) is constructed on the top of the
model by incorporating Maximum a Posteriori (MAP) decision rule.
The classifier function assigns a class label $\widehat{y}=cL_{j}^{i}$
for $i \in \{1,2,3\}$ as shown in Eq. \ref{classifier}.

\begin{flalign}
\label{classifier}
\widehat{y}= \argmax_{i\in \lbrace 1,2,3 \rbrace} p(cL_{j}^{i}) \prod_{k=1}^{\lvert \varsigma \rvert} p(\varsigma^{k} | cL_{j}^{i}) \rbrace
\end{flalign}

It is likely that an incoming VM $V_j \in G^v(N^v,E^v)$ classified into class $cL_{j}^{i=1}$, $cL_{j}^{i=1}$, and $cL_{j}^{i=1}$ with equal probability. $p(cL_{j}^{i=1})$=$p(cL_{j}^{i=2})$=$p(cL_{j}^{i=3})$=$\frac{1}{\#~of~classes}$=$\frac{1}{3}$=$0.33$.

\subsection{Iterative reinforcement learning}

The Reinforcement Learning (RL) is a powerful tool of artificial
intelligence, where the system often called as an agent gradually
trains itself to act in a manner that yields the incremental reward
by means of interacting with the given environment. The RL differs
from the established notion of supervised machine learning, where
the system learns under the supervision of labelled samples. The RL agent learns in real-time from the given state of the
environment to improve the overall outcome. The RL problem is
usually considered as a set of sub-optimal actions that together
form a Markov Decision Process (MDP). The RL has to walk a tight
rope by balancing a trade-off between the exploitation and
exploration of the environment. In order to maintain the system
performance, the RL agent explores the environment (Cloud data centers) and learns about the existing utilized resources. Subsequently, RL agent exploits the environment to determine the effective action (VNE) that results into positive rewards.
To be specific, the
RL agent simultaneously explores the environment and exploits the alternative actions that gradually yields the incremental rewards in subsequent actions.
The powerful feature of RL is that the learning process looks to
achieve a global optimum solution instead of looking for the
immediate sub-optimum solution.

Let us consider that RL agent is at the specific state $s_t$ of a
cloud environment. The agent takes action $a_t$ based on the
learning, and in response, the cloud environment provides certain
reward $r_t$ and changes to the new state $\overline{s}_t$. Let us
denote the aforementioned process as
$<s_t,a_t,r_t,\overline{s}_t>$. In the context of the proposed VNE
in the cloud environment, the action is defined as the selection
of SNs for the given configuration of the incoming VMs and the
reward is the quality of the embedding. In VNE, the above
described process continues and takes finite number of sequences
of states, actions and rewards. The entire MDP process can be
described as
$\{<s_0,a_0,r_0,\overline{s_1}>,...,<s_t,a_t,r_t,\overline{s_{t+1}}>,...,<s_{n},a_n,r_n,\overline{s_{n+1}}>\}$.

%

Here, $<s_t,a_t,r_t,\overline{s_{t+1}}>$ describe the $t^{th}$
instance of the action taken by the RL agent, where $s_t$
represents the $t^{th}$ state of the cloud environment, $a_t$
represents the $t^{th}$ action taken by the RL agent, $r_t$
represents the $t^{th}$ reward provided by the environment, and
$s_{t+1}$ represents the next $(t+1)^{th}$ state of the
environment. The State-action-reward-state-action (SARSA)
algorithm is applied to learn the aforementioned Markov decision
process. The goal is to update the policy, which is referred to as
the optimum state-action value function $Q(s_t, a_t)$ based on the
action taken by the SARSA agent in a cloud environment as shown in
Eq. \ref{SARSA}.

\begin{flalign}
\label{SARSA}
Q(s_t, a_t) \leftarrow Q(s_t, a_t) + \alpha \left[r_t + \gamma Q(s_{t+1}, a_{t+1})-Q(s_t, a_t)\right]
\end{flalign}

\begin{figure}
    \centering
    \includegraphics[width=0.7\linewidth]{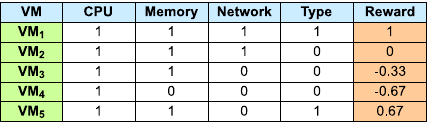}
    \caption{The example calculation of environment reward.}
    \label{fig:rewardexample}
\end{figure}

Here, $Q(s_t, a_t)$ represents the value function for the state
$s_t$ and action $a_t$. The $\alpha \in [0,1]$ represents the
learning rate and $\gamma\in [0,1]$ represents the discount
factor. Eq. \ref{SARSA} represents the possible reward received in
the subsequent step for taking action $a_{t}$ in state $s_{t}$
along with the discounted future reward received in the state
$s_{t+1}$ for action $a_{t+1}$. In order to maximize the objective
function defined in Eq. \ref{eq:objective}, the SARSA agent learns
and takes the actions in a manner that maximizes the rewards from
the environment. The reward calculation can be described as
follow. Upon the classification of VM type, SARSA agent selects
the set of SNs that not only satisfy the resource requirement of
the VMs in a given VN request but at the same time schedule them
onto the appropriate SN type. For example, the CPU intensive VM
must be scheduled onto the CPU rich SN along with satisfying the
other resource requirements. Based on the action referred to as
the VNE by the RL agent, a unique value between -1 to +1 is
assigned by the cloud environment indicating the reward to the
action. The more the positive reward the better the action was taken
by the SARSA agent. The reward is calculated as follow. First, the
reward is equally distributed between resource requirement
satisfaction and type selection. To be specific, to satisfy the VM
type, the agent receives a +0.5 reward. Similarly, for satisfying
the resource requirement irrespective of the type, the +0.5 reward
is awarded. Again, the +0.5 resource reward is equally divided
among each resource type such as CPU, memory, and network. If any
one of the resource requirements is not satisfied -0.17 reward is
awarded. The example of rewards under different conditions is
illustrated in Fig. \ref{fig:rewardexample}.

The reward $\mathbb{R}$ of VN $(G^{v}(N^{v},E^{v}))$ can be formally defined as follow.

\begin{align}
\label{reward}
\mathbb{R}(G^{v}(N^{v},E^{v}))= \sum_{V_j \in G^{v}(N^{v},E^{v})} \sum_{x \in \{CPU, Memory, Network\}} \mathbb{R}(x) \times \mathbb{H}
\end{align}

Here, $\mathbb{R}$ represents the individual reward received on
resource type $x$, and $\mathbb{H}$ represents the decision
variable. The $\mathbb{H}$ takes either -1 or +1 value. The
$\mathbb{H}$ can be defined formally as follow.

\begin{equation*}
\mathbb{H} = \begin{cases}
1 & A_{i}^{x} = D_{i}^{x}\\
-1 & A_{i}^{x} \ne D_{i}^{x}
\end{cases}
\end{equation*}

The $\mathbb{R}(x)$ can take value either $0$ or $0.17$.

\section{Performance Evaluation and results} \label{sec:perfEvaluation}
In the current section, the performance of the proposed MUVINE
scheme is evaluated using the light-weight Java-based discrete event simulator, which is used to generate the substrate network and virtual network requests. Various distribution functions are implemented such as Poisson distribution to simulate the VN request arrival, random distribution to distribute the resources to virtual nodes, virtual links, and substrate network, etc. The performance of the proposed binary VN classifier is
compared with the recent admission control algorithm RNN\_VNE
\cite{blenk2016boost}. The simulation parameters for the SNs, VNs,
and VMs are listed in Table \ref{SimulationforPM}, Table
\ref{SimulationforVN}, and Table \ref{SimulationforVM},
respectively.

\subsection{Simulation Setup}\label{sec:perfEvaluation:simSetup}
In this section, the simulation setup for SNs, VNs, and VMs is described in detail. The simulation configuration for the Substrate and Virtual Networks with their resource configuration is taken to suit this small scale simulation environment and for better understanding of the performance evaluation results. However, simulation configuraiton can be extended without any further modification to the simulation environment. Similar simulation configurations were used in our previous works LVRM \cite{sahoo2018lvrm} and DIVYNE \cite{dehury2019dyvine} in order to rigorously evaluate the VNE embedding schemes. During the simulation, one service provider is considered equipped with 100 SNs connected through links that are generated randomly. For the random links generation, a probability value is set to 0.6, which also acts as the connectivity probability of two SNs. Random distribution is used to allocate the resources to SNs. A randomized function is used to assign the available number of CPUs capacities to each SN ranging from 16 to 32 CPUs. Similarly, the storage and memory capacity are randomly distributed in between 500 GB through 1000 GB, and 20000 MB and 50000 MB, respectively. The available bandwidth between the pairs of SNs is randomly distributed between 1000 Bps and 10000 Bps. The Bps refers to Bytes per second.

In order to generate VN requests, each VN is equipped with VMs in the range of 2 to 10. It is assumed that VN request arrival follows the Poisson distribution with the mean of 5 requests per one hundred time units, and the lifetime of each request follows the exponential distribution with an average lifetime of five hundred time units. It is to note that in a few cases the unexpired lifetime of requests is also considered. The maximum number of virtual links for each request is 45. The number of virtual links is decided by the link probability of 0.6, which also represents the connectivity probability of two VMs. The resource demand for each request is randomly distributed. The number of CPUs demand ranges from 1 through 4. Similarly, the virtual links range from 100 Bps through 500 Bps. The required storage and memory for each VM range from 500 GB to 1000 GB, and 8000 MB to 10000 MB, respectively. The aforementioned simulation setup is repeated several times to ensure sufficient variations in the number of VN requests arrival, types of VN request arrival, amount of resource allocations to VMs etc. A $t-$test is performed on simulation generated dataset to determine the significant difference between random samples. The $P < 0.05$ is obtained, which indicates the statistical significance of the simulation dataset. 

In the proposed MUVINE scheme, the Scikit-learn \cite{scikit-learn} open source machine learning library is used to implement the supervised prediction algorithms such as SVM, RBF, and MLC; whereas the implementation of RL is carried out using the TensorFlow software library \cite{abadi1603tensorflow}. In the proposed MUVINE scheme, the SVM, RBF, and MLC are trained on the 66\% samples randomly selected from the simulated dataset called as training dataset. On the other hand, the same prediction algorithms are evaluated on remaining 33\% of samples also called as testing dataset. The training dataset is comprised of samples that include the features and corresponding real labels. For instance, MUVINE scheme implements the SVM to accept or reject the VN request. Here, each training sample of SVM is comprised of values for VN features described in Section \ref{VN_feature_set} along with a unique label of "accept" or "reject". On the contrary, each test sample is comprised of only values for the VN features and corresponding label is predicted by the SVM. Typically, a loss function is employed to evaluate the performance of supervised algorithms by measuring the distance between the real and predicted label. Contrary to the supervised learning, RL has features without real label of the samples. To be specific, RL does not have dataset, but it works on the given environment. In the current VNE problem, the RL agent works on the cloud environment that includes the VM type, SN type, CPU and memory demand of each VM, available CPU, memory, and bandwidth of each SN etc. The RL explores the states and exploits the all possible actions to predict the most suitable action (i.e., VNE embedding) that results in the improved performance of an evaluation metric also called as reward as defined in Eq. \ref{reward}. Considering the aforementioned simulation setup, the simulation results are obtained as follows.


\begin{table}
    \caption{The simulation setup for Substrate Nodes (SNs)}
    \centering
    \begin{tabular}{|c|c|}
        \hline
        \textbf{Parameters} & \textbf{Value} \\
        \hline
        \multicolumn{2}{|c|}{Total number of SNs = 100} \\
        \hline
        \multicolumn{2}{|c|}{Overall SN resource utilization} \\
        \hline
        Minimum utilization & 80\% \\
        \hline
        Maximum utilization & 95\% \\
        \hline
        \multicolumn{2}{|c|}{CPU capacity of each SN} \\
        \hline
        Minimum CPU capacity & 16 \\
        \hline
        Maximum CPU capacity & 32 \\
        \hline
        \multicolumn{2}{|c|}{Memory capacity of each SN (in MB)} \\
        \hline
        Minimum memory capacity & 2000 \\
        \hline
        Maximum memory capacity & 5000 \\
        \hline
    \end{tabular}
    \label{SimulationforPM}
\end{table}

\begin{table}
    \caption{The simulation setup for Virtual Networks (VNs)}
    \centering
    \begin{tabular}{|c|c|}
        \hline
        \textbf{Parameters} & \textbf{Value} \\
        \hline
        \multicolumn{2}{|c|}{Total number of VN requests = 15000} \\
        \hline
        \multicolumn{2}{|c|}{Size of the VN} \\
        \hline
        Minimum number of VMs & 2 \\
        \hline
        Maximum number of VMs & 10 \\
        \hline
        \multicolumn{2}{|c|}{Other hyper parameters} \\
        \hline
        Start time & 0.05 seconds \\
        \hline
        Mean life time & 50 seconds \\
        \hline
    \end{tabular}
    \label{SimulationforVN}
\end{table}

\begin{table}
    \caption{The simulation setup for Virtual Machines (VMs)}
    \centering
    \begin{tabular}{|c|c|}
        \hline
        \textbf{Parameters} & \textbf{Value} \\
        \hline
        \multicolumn{2}{|c|}{Required memory (in MB) of VM} \\
        \hline
        Minimum required memory & 500 \\
        \hline
        Maximum required memory & 4096 \\
        \hline
        \multicolumn{2}{|c|}{Required number of CPUs by VM} \\
        \hline
        Minimum number of CPUs & 1 \\
        \hline
        Maximum number of CPUs & 4 \\
        \hline
        \multicolumn{2}{|c|}{Actual memory usage (in \%) by VM} \\
        \hline
        Minimum memory usage & 30\% \\
        \hline
        Maximum memory usage & 99\% \\
        \hline
        \multicolumn{2}{|c|}{Actual CPU usage (in \%) by VM} \\
        \hline
        Minimum CPU usage & 30\% \\
        \hline
        Maximum CPU usage & 99\% \\
        \hline
        \multicolumn{2}{|c|}{Scheduling class of VM} \\
        \hline
        Minimum scheduling class & 1 \\
        \hline
        Maximum scheduling class & 3 \\
        \hline
        \multicolumn{2}{|c|}{Priority class of VM} \\
        \hline
        Minimum priority class & 1 \\
        \hline
        Maximum priority class & 3 \\
        \hline
    \end{tabular}
    \label{SimulationforVM}
\end{table}

%
%

\subsection{Simulation Results}
The simulation of MUVINE scheme is carried out in three stages. Each prediction stage is simulated independently and the simulation results are reported. The simulation results followed by the discussions for each prediction stage are reported as follow.

\subsubsection{VN selection accuracy}
Upon VN request arrival, the user-defined and CSP-defined parameters are analyzed using the proposed binary SVM classifier denoted as MUVINE\_SVM to predict the acceptability of a VN request. For the detailed evaluation, the predictability of MUVINE\_SVM is analyzed multi-faceted across the time-domain and across the VN request arrival etc. The prediction outcome of MUVINE\_SVM is then compared with the recent Recurrent Neural Network (RNN) based cloud admission control algorithm denoted as RNN\_VNE \cite{blenk2016boost}.

\begin{figure}
    \centering
    \subfigure[]{\label{fig:a}\includegraphics[width=83mm]{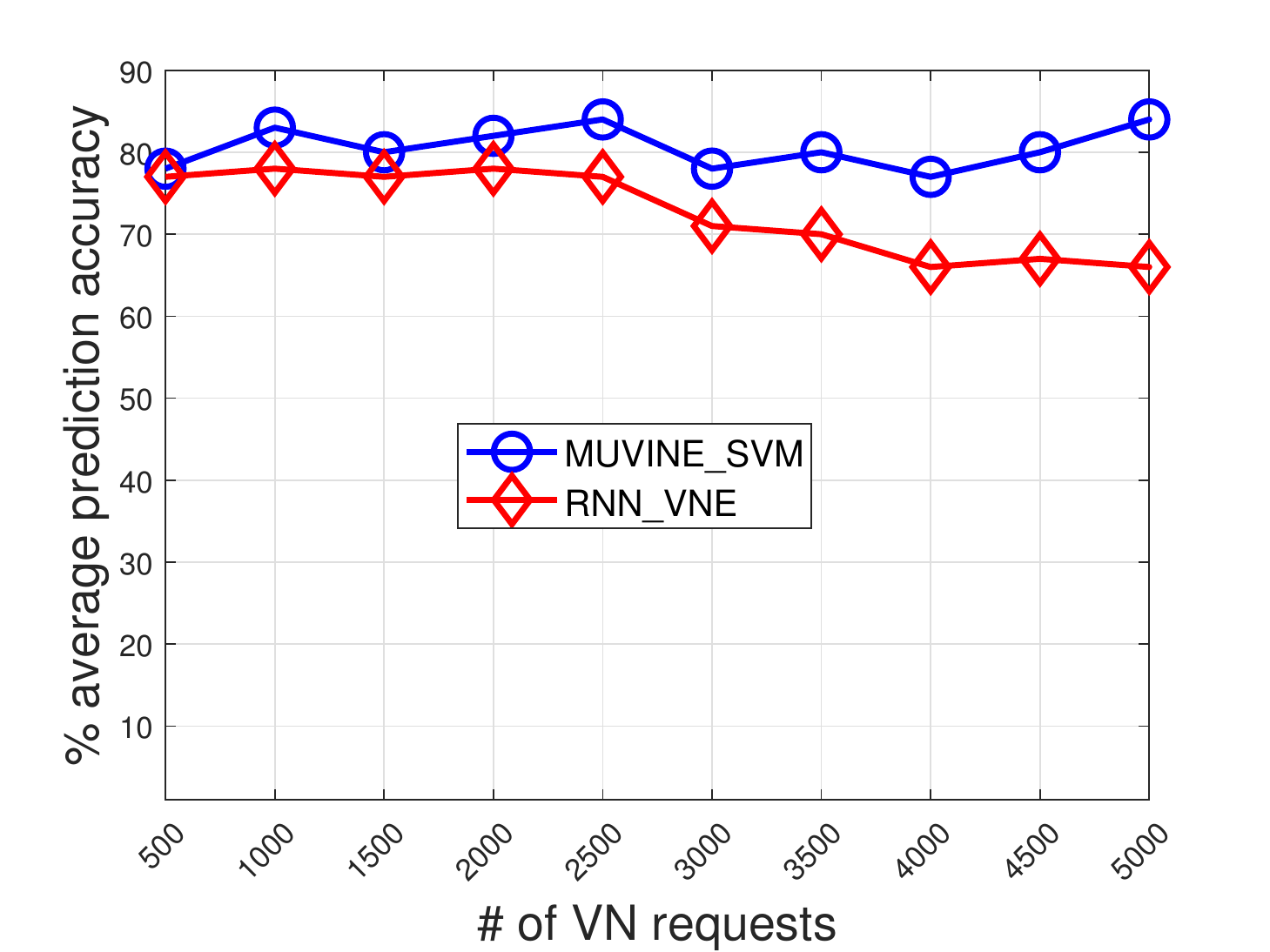}}
    \hspace*{-0.9em}
    \subfigure[]{\label{fig:b}\includegraphics[width=83mm]{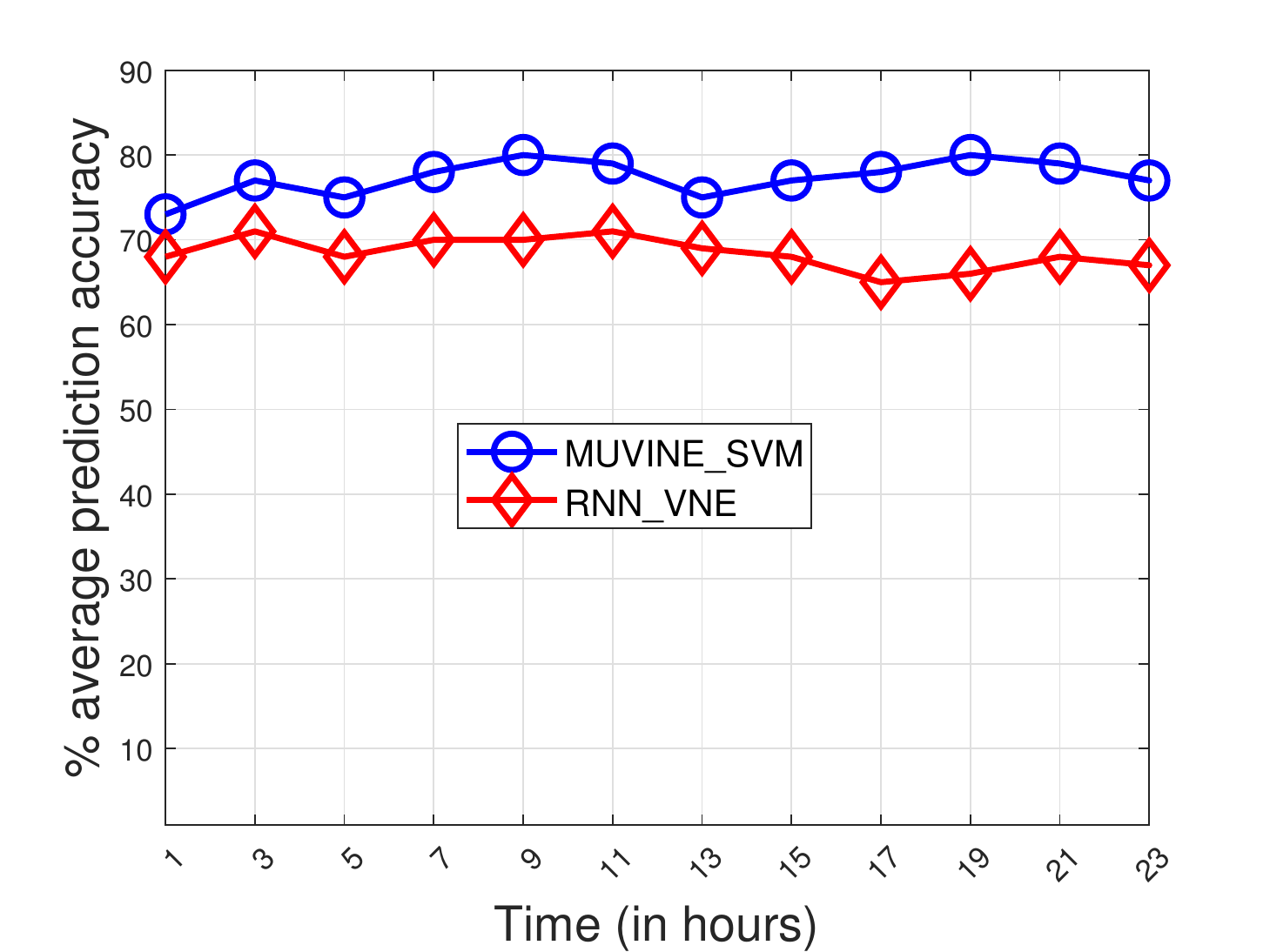}}
    \caption{The comparison of average prediction accuracy (in percentage) with respect to (a). number of VN requests, and (b). time (in hours).}
    \label{fig:vnpredrequests_time}
\end{figure}


Fig. \ref{fig:vnpredrequests_time}a shows the average prediction accuracy (in percentage) of MUVINE\_SVM for different number of VN requests. The 5000 test samples are evaluated at the interval of 500 VN requests and corresponding observed prediction accuracy is reported. As shown in Fig. \ref{fig:vnpredrequests_time}a, the average prediction accuracy (in percentage) of the proposed MUVINE\_SVM is consistent and as high as 84\%. Contrary to the RNN\_VNE, the MUVINE\_SVM achieves average prediction accuracy improvement of 8\%. 


Fig. \ref{fig:vnpredrequests_time}b shows the effectiveness of the MUVINE\_SVM with respect to the time. The performance of MUVINE\_SVM is evaluated for the 24 hours duration and later compared with the RNN\_VNE \cite{blenk2016boost}. As shown in Fig. \ref{fig:vnpredrequests_time}b, MUVINE\_SVM consistently outputs improved prediction accuracy irrespective of the VN requests' arrival time. The MUVINE\_SVM is robust against the highly fluctuating nature of VN workload arrival. Similar to our previous observation, the proposed MUVINE\_SVM shows at least 8.8\% higher average prediction accuracy to that of RNN\_VNE \cite{blenk2016boost} in the time domain.

\begin{figure}
    \centering
    \includegraphics[width=0.8\linewidth]{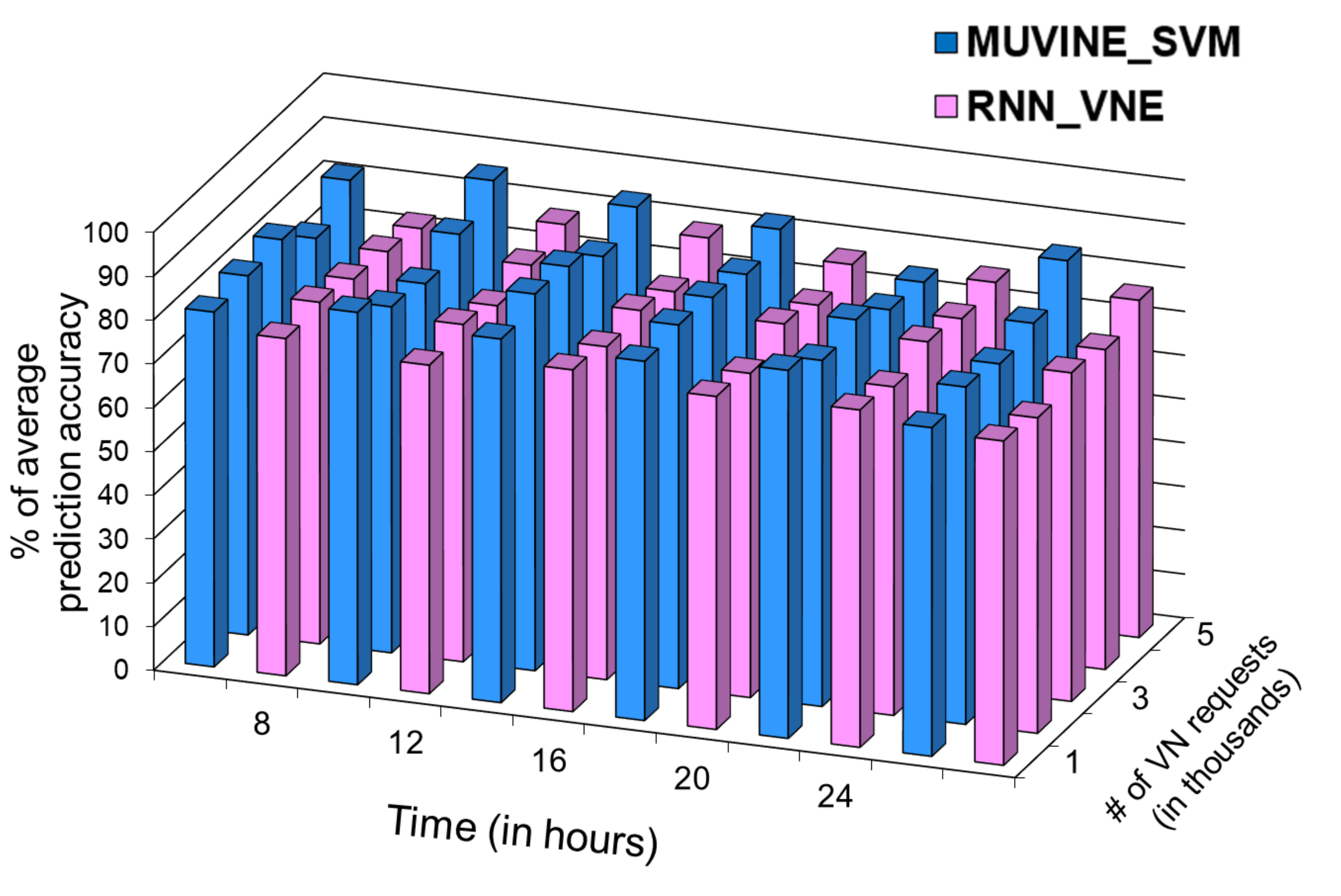}
    \caption{The prediction accuracy of VN requests}
    \label{fig:vn3dpred}
\end{figure}

For the detailed and extensive performance evaluation, the MUVINE\_SVM is simultaneously evaluated against VN request arrival time and number of VN requests together. Fig. \ref{fig:vn3dpred} shows the outcome of the MUVINE\_SVM and RNN\_SVM. As shown in Fig. \ref{fig:vn3dpred}, the performance of MUVINE\_SVM is consistent and it achieves higher prediction accuracy irrespective of the number of VN requests and their arrival time. The MUVINE\_SVM outperforms RNN\_VNE \cite{blenk2016boost} with as much as 6.8\% higher prediction accuracy across the VN request arrival time and number of VN requests. The possible explanations of SVM outperforming over RNN\_VNE are as follow. The MUVINE\_SVM is implemented on the top of the carefully selected performance centric VN features discussed in \ref{VN_feature_set}, which has empowered the MUVINE\_SVM to better distinguish the acceptability of an incoming VN request as compared to the RNN\_VNE \cite{blenk2016boost}. Contrary to RNN\_VNE, the MUVINE\_SVM scales relatively well to the high dimensional data with numerous features, which makes it more suitable to deal with numerous VN requests of cloud environment. When the underlying training and test data can be separated with the hyperplane, the SVM provides a robust outcome with relatively less training and testing time. Considering the VN admission decision as the first stage of the proposed Multi-stage MUVINE scheme, the possible delay leads to the overall delay in VNE. Therefore, the VN predictor is designed with the simpler yet robust alternative to RNN.


\begin{figure*}
	\centering
	\subfigure[CPU intensive VM class prediction]{\label{fig:a}\includegraphics[width=82mm]{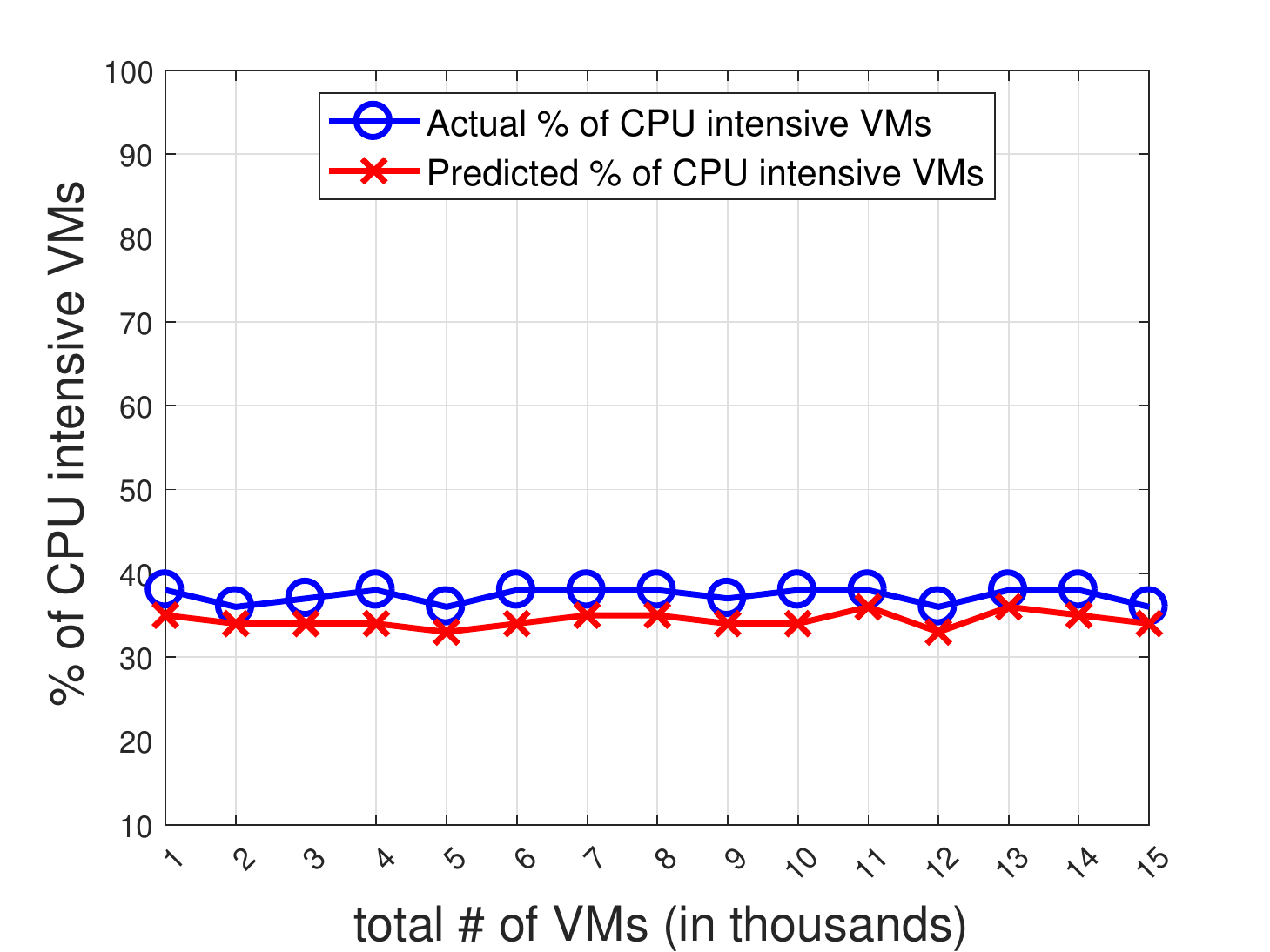}}\hspace*{-1.7em} \\
	\subfigure[Memory intensive VM class prediction]{\label{fig:b}\includegraphics[width=82mm]{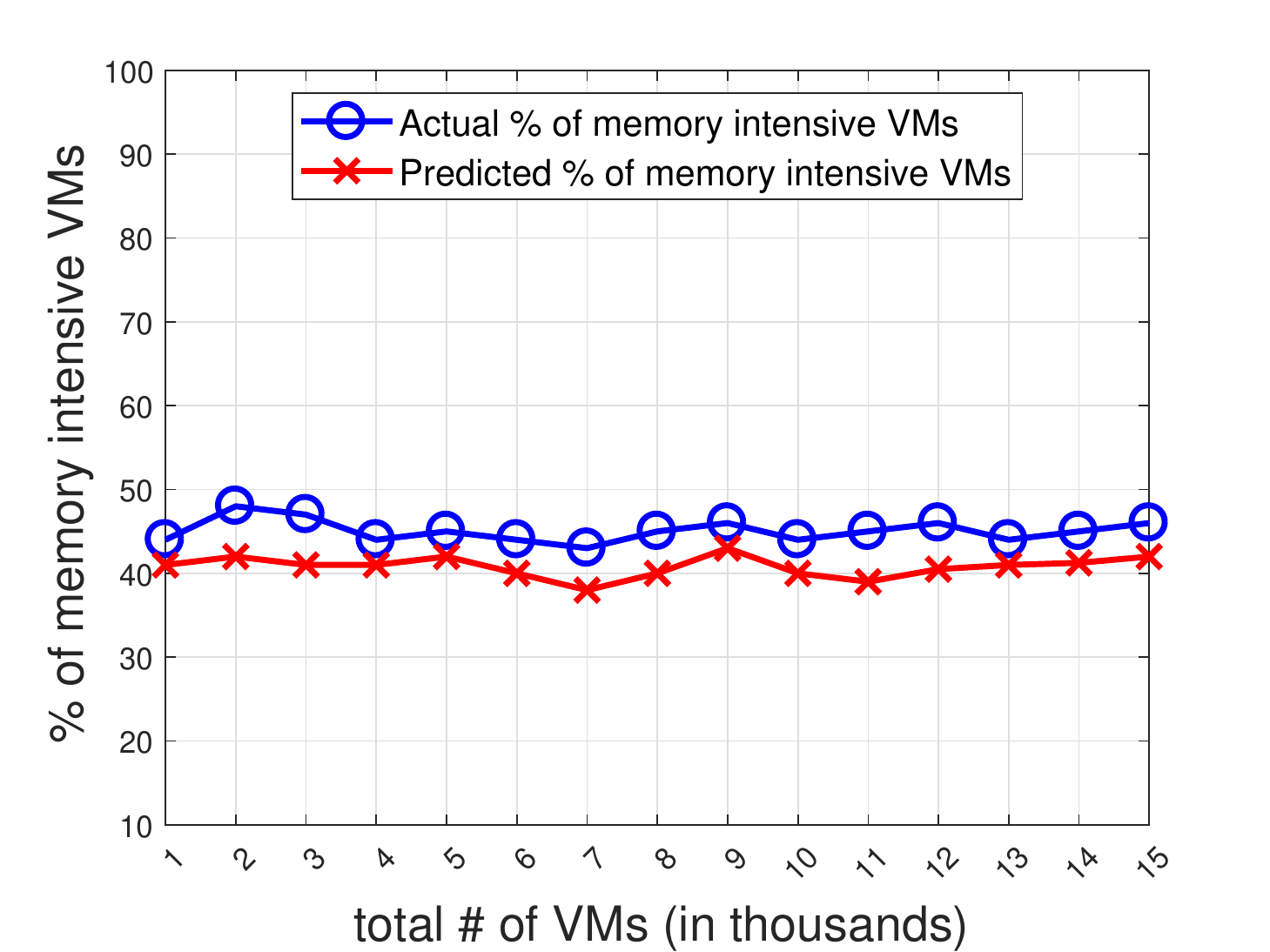}}\hspace*{-1.7em}
	\subfigure[GPU intensive VM class prediction]{\label{fig:c}\includegraphics[width=82mm]{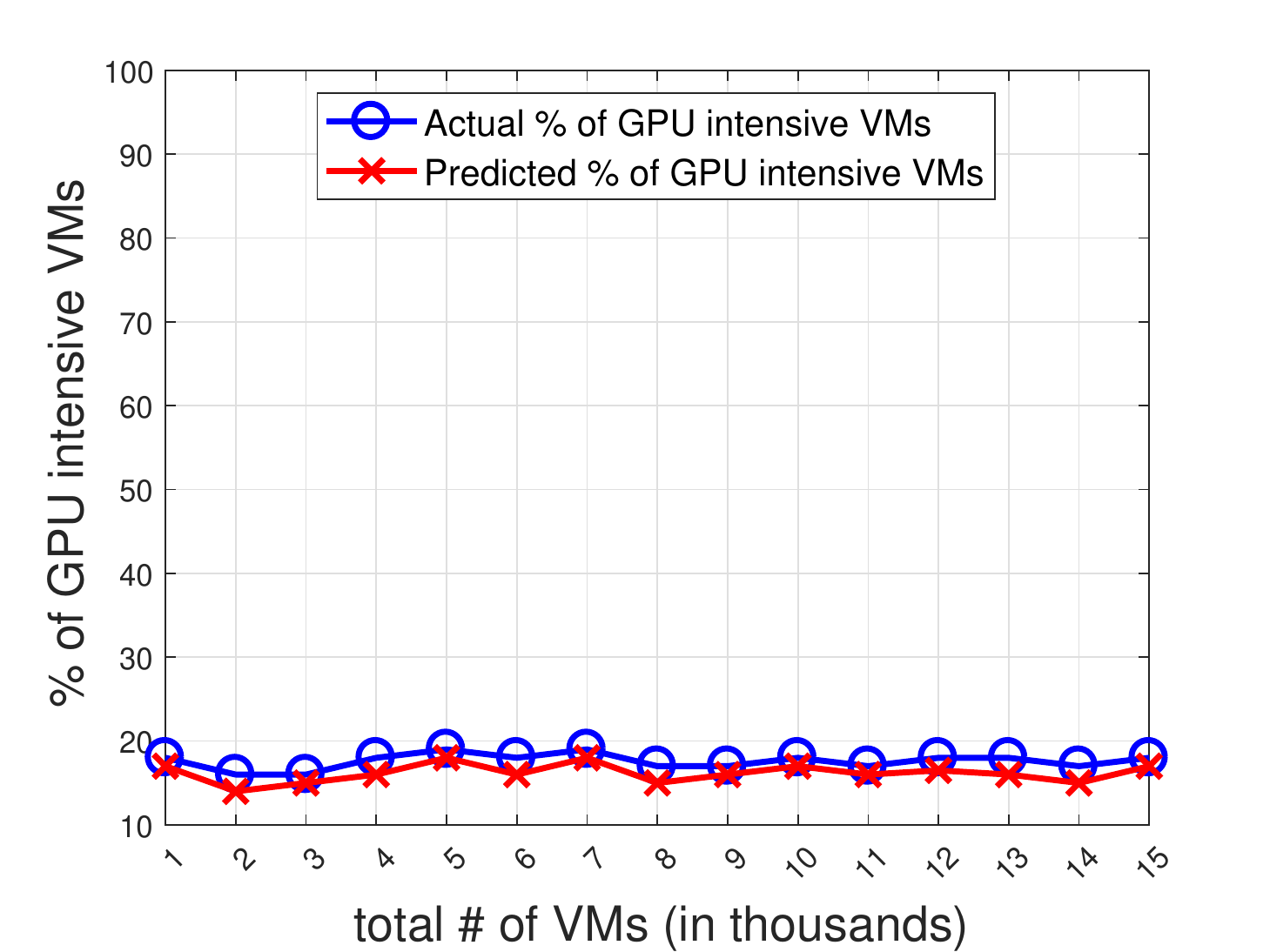}}
	\caption{The performance of Maximum Likelihood Classifier (MLC) for (a). CPU intensive VM class prediction, (b). Memory intensive VM class prediction, (c). GPU intensive VM class prediction.}
	\label{fig:CPU_Mem_GPU_NoOfVMs}
\end{figure*}

\subsubsection{VM classification accuracy}
Upon acceptance of the new VN request, the proposed MUVINE scheme predicts the "end time" and "resource utilization" of the corresponding VMs. The "end time" and "resource utilization" are the derived features as discussed in Section \ref{sec:2ndStage}. The core and derived features are further used in order to classify the VMs into 3 different categories. The performance of the second stage VM class prediction is presented in Fig. \ref{fig:CPU_Mem_GPU_NoOfVMs}. Fig. \ref{fig:CPU_Mem_GPU_NoOfVMs}a shows the accuracy of the prediction of CPU intensive VM class by comparing the prediction result with that of the actual VM class. The information of actual VM class is obtained from the CSP post-execution of VN request. The x-axis in Fig. \ref{fig:CPU_Mem_GPU_NoOfVMs}a represents the total number of VMs ranging from $1000$ to $15000$; whereas the y-axis represents the percentage of the CPU intensive VMs. The analysis of the simulation dataset reveals that nearly $30\%$ to $40\%$ VMs are the CPU intensive VMs and the rest are either memory intensive or GPU intensive. The simulation result of Fig. \ref{fig:CPU_Mem_GPU_NoOfVMs}a shows that the prediction result is quite close to the actual one. From $1000$ VMs, it is predicted that approximately $35\%$ i.e. $350$ VMs belong to CPU intensive class with an error of approximately $7\% $. However, the prediction result marginally improves with the increase in the number of VMs from $1000$ to $15000$. The prediction result shows that approximately $34\%$ VMs are classified as CPU intensive with an error of around $2\%$, which shows the improved classification accuracy.



On the line of the prediction result of CPU intensive VMs, Fig. \ref{fig:CPU_Mem_GPU_NoOfVMs}b shows the performance for the memory-intensive VMs. The ground truth from the simulation result reveals that the actual number of memory-intensive VMs ranges between $40\%$ to $48\%$, which is slighly higher than that of the CPU intensive VMs in the simulation data. However, the prediction error ranges between $7\%$ and $3\%$. Out of $1000$ VMs, nearly $41.5\%$ of the VMs are predicted memory intensive in contrast to $44\%$ actual memory-intensive VMs. Similarly, in the case of $15000$ VMs, $42\%$ of the VMs are predicted memory intensive VMs; whereas the number of actual memory-intensive VMs is $45.5\%$. Similar prediction result is observed in the prediction of GPU intensive VMs as in Fig. \ref{fig:CPU_Mem_GPU_NoOfVMs}c. The ground truth from the simulation result reveals that the percentage of the actual GPU intensive VMs ranges between $15\%$ and $19\%$. However, predicted percentage of the GPU intensive VMs ranges between $13.5\%$ and $18\%$ with a maximum error $6\%$ and minimum error $1.5\%$. 

The noticeable performance of the proposed multi-class Maximum Likelihood Classifier (MLC) can be explained as follows. The proposed MLC is trained on the aggregate features instead of the core one. The inclusion of the derived features makes the underlying aggregate feature set more robust, which empowers the MLC to look for the complex relationship among the VM types and their corresponding feature values. It is to be noted that the proposed MLC independently calculates the likelihood of features for a given VM type. The MLC performs well due to its inherent assumption of conditional independence among the features, which is highly relevant in the cloud environment. For instance, CPU, memory, and GPU intensive VMs have resource requirement that are very specific to the CPU, memory, and GPU, respectively. Therefore, the concerned features dominate more and have least direct relation with other features. The MLC takes the advantage of this cloud property and efficiently predicts the VM classes. The simple to implement yet robust MLC is chosen, considering the requirement of the real-time classification of VMs in the cloud.

\subsubsection{VNE evaluation}
Upon classifying the VMs, the proposed MUVINE\_RL
selects the suitable SNs for the embedding purpose. The resource
utilization is one of the foremost parameters to evaluate the
performance of VNE. The better the resource utilization, the
better the VNE. Here, two types of resources are considered for
the evaluation such as CPU and Memory. The performance
evaluation of MUVINE\_RL is carried out with respect to the time
domain and number of VMs requests arrival. The performance of the
MUVINE\_RL is compared with the two traditional state-of-the-art
techniques such as Game\_Alloc \cite{Wei2018Imperfect} and LVRM
\cite{sahoo2018lvrm}, and three AI-based techniques such as Auto\_RP
\cite{GHOBAEIARANI2018191}, RL\_VNE \cite{yao2018novel}, and RNN\_VNE \cite{blenk2016boost}.

\begin{figure}
    \centering
    \subfigure[]{\label{fig:a}\includegraphics[width=83mm]{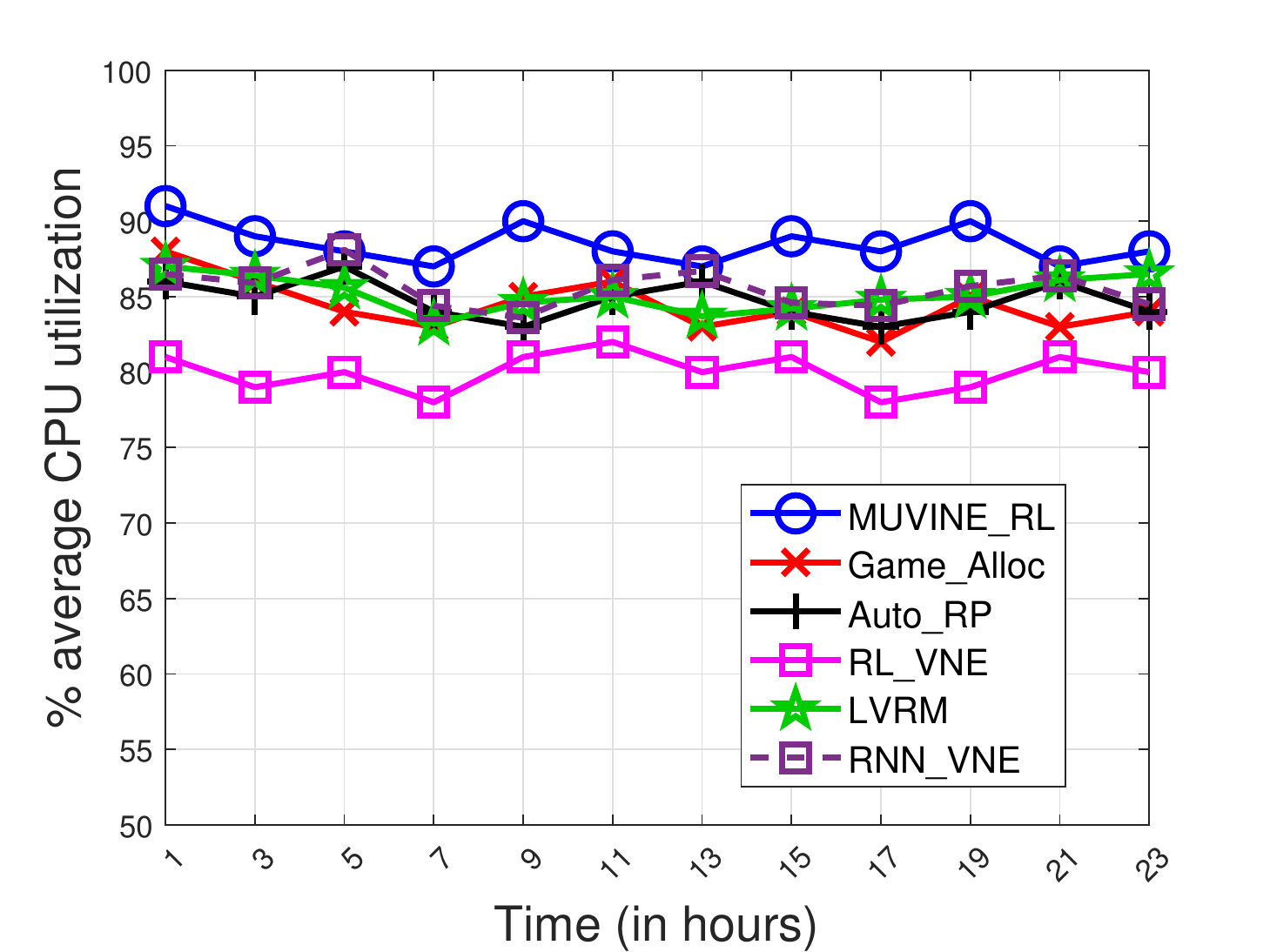}}
    \hspace*{-0.9em}
    \subfigure[]{\label{fig:b}\includegraphics[width=83mm]{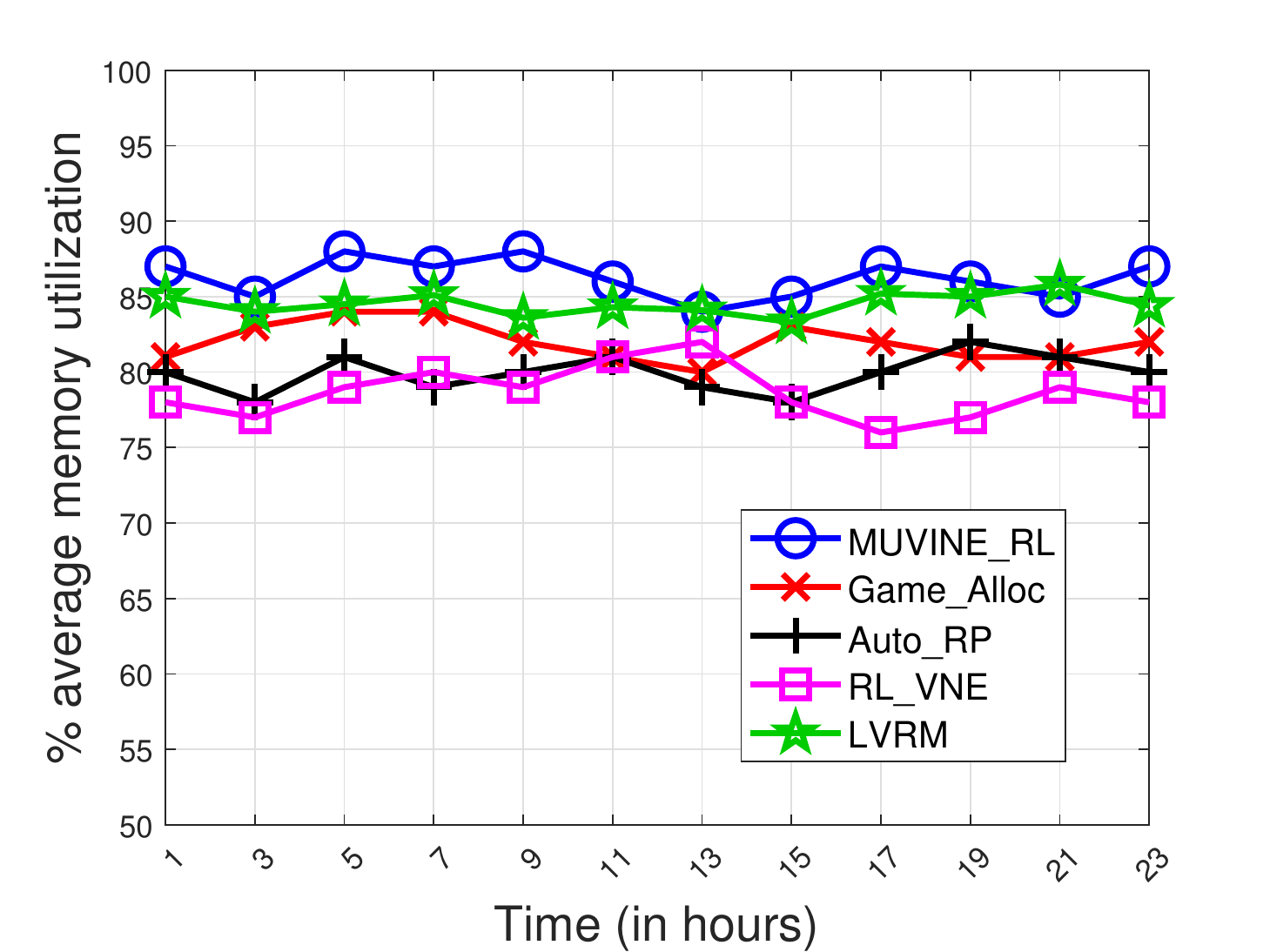}}
    \caption{The comparison of average resource utilization (in percentage) with respect to time domain for (a). CPU, (b). Memory.}
    \label{fig:3rdstagecpumemoryrutime}
\end{figure}

\begin{figure}
    \centering
    \subfigure[]{\label{fig:a}\includegraphics[width=83mm]{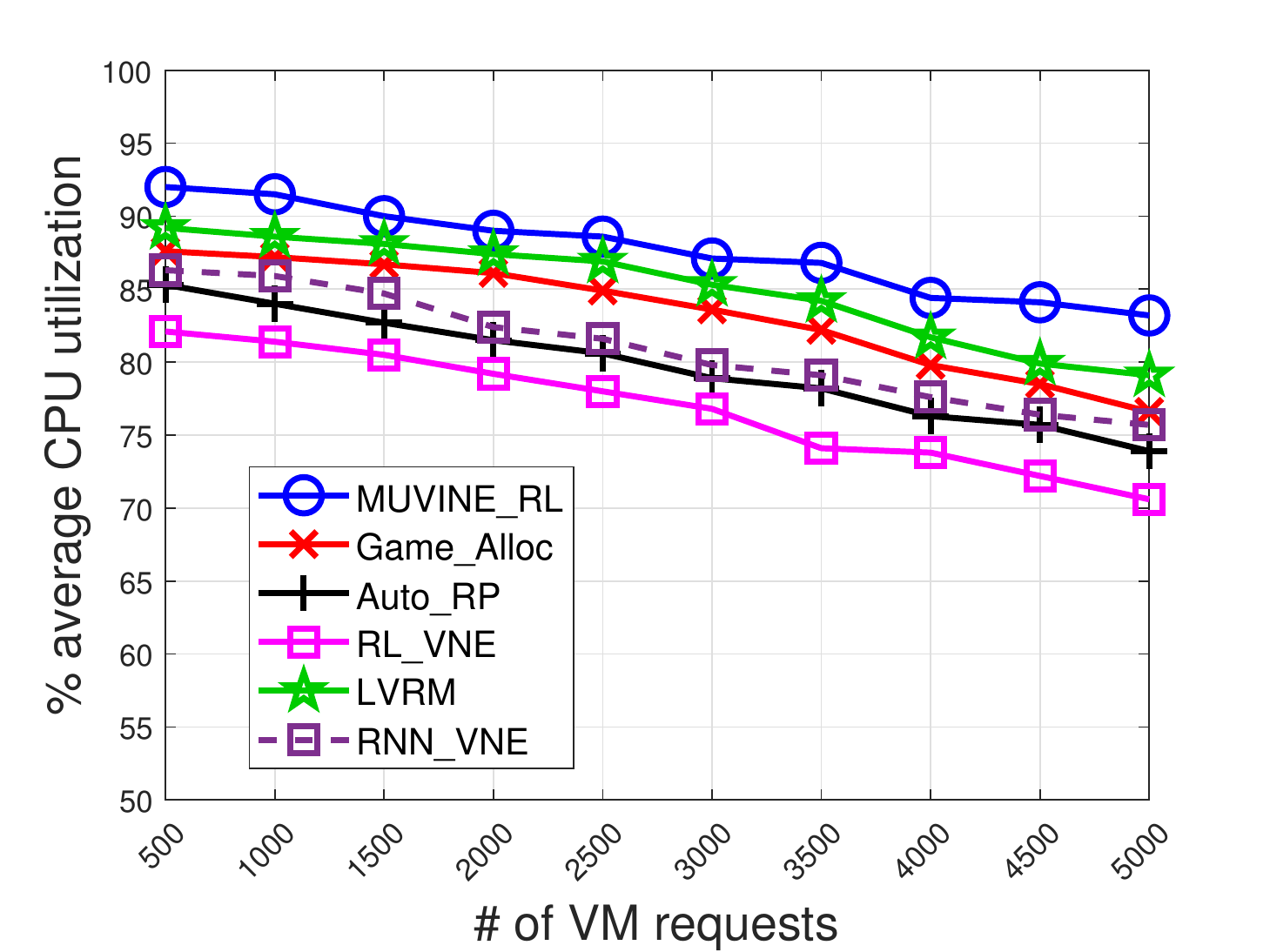}}
    \hspace*{-0.9em}
    \subfigure[]{\label{fig:b}\includegraphics[width=83mm]{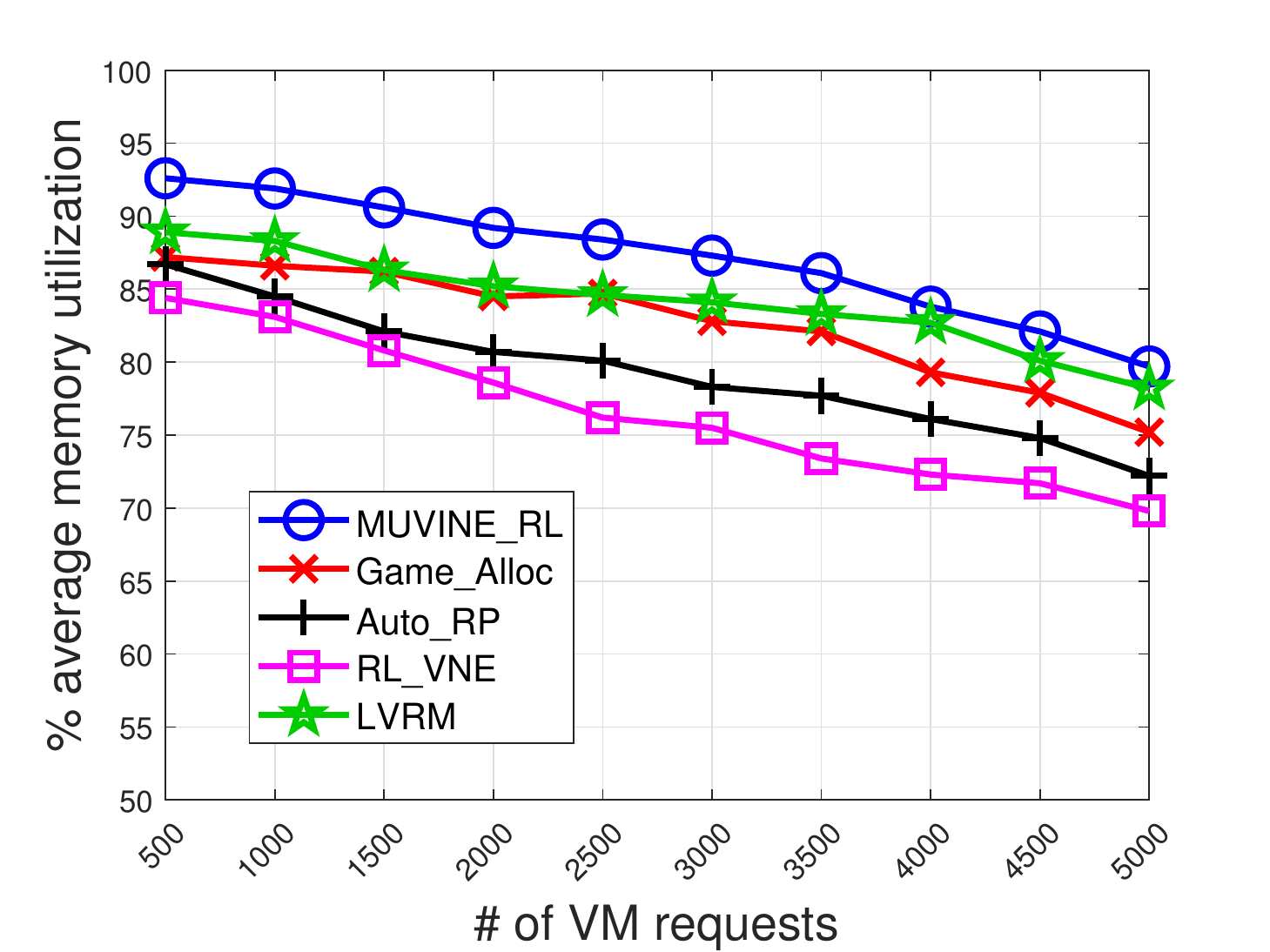}}
    \caption{The comparison of average resource utilization (in percentage) with respect to requests for (a). CPU, (b). Memory.}
    \label{fig:3rdstagecpumemoryrurequests}
\end{figure}

Fig. \ref{fig:3rdstagecpumemoryrutime}a and Fig. \ref{fig:3rdstagecpumemoryrutime}b shows the percentage average CPU and memory utilization (in percentage) with respect to the time domain, respectively. With each passing hour, the performance of each VNE scheme is monitored and the CPU and memory utilization are recorded. From Fig. \ref{fig:3rdstagecpumemoryrutime}a and Fig. \ref{fig:3rdstagecpumemoryrutime}b, it is clear that all VNE schemes are consistent and they maintain the amount of average CPU and memory utilization in the time domain. However, the average CPU and memory utilization of the Game\_Alloc \cite{Wei2018Imperfect}, LVRM \cite{sahoo2018lvrm}, Auto\_RP \cite{GHOBAEIARANI2018191}, RL\_VNE \cite{yao2018novel}, and RNN\_VNE \cite{blenk2016boost} is less as compared to that of MUVINE\_RL. The percentage of average utilization of Game\_Alloc \cite{Wei2018Imperfect}, LVRM \cite{sahoo2018lvrm}, Auto\_RP \cite{GHOBAEIARANI2018191}, RL\_VNE \cite{yao2018novel}, RNN\_VNE \cite{blenk2016boost}, and MUVINE\_RL is 83.3\%, 85.0\%, 79.7\%, 76.8\%, 85.3\%, and 87.6\%, respectively. Besides, the MUVINE\_RL achieves improved consistency with standard deviation in CPU utilization as low as 2.93 compared to that of 3.7, 3.5, 3.56, 3.8, and 3.1 of Game\_Alloc \cite{Wei2018Imperfect}, LVRM \cite{sahoo2018lvrm}, Auto\_RP \cite{GHOBAEIARANI2018191}, and RL\_VNE \cite{yao2018novel}, RNN\_VNE \cite{blenk2016boost}, respectively. 

\begin{figure}
    \centering
    \includegraphics[width=0.6\linewidth]{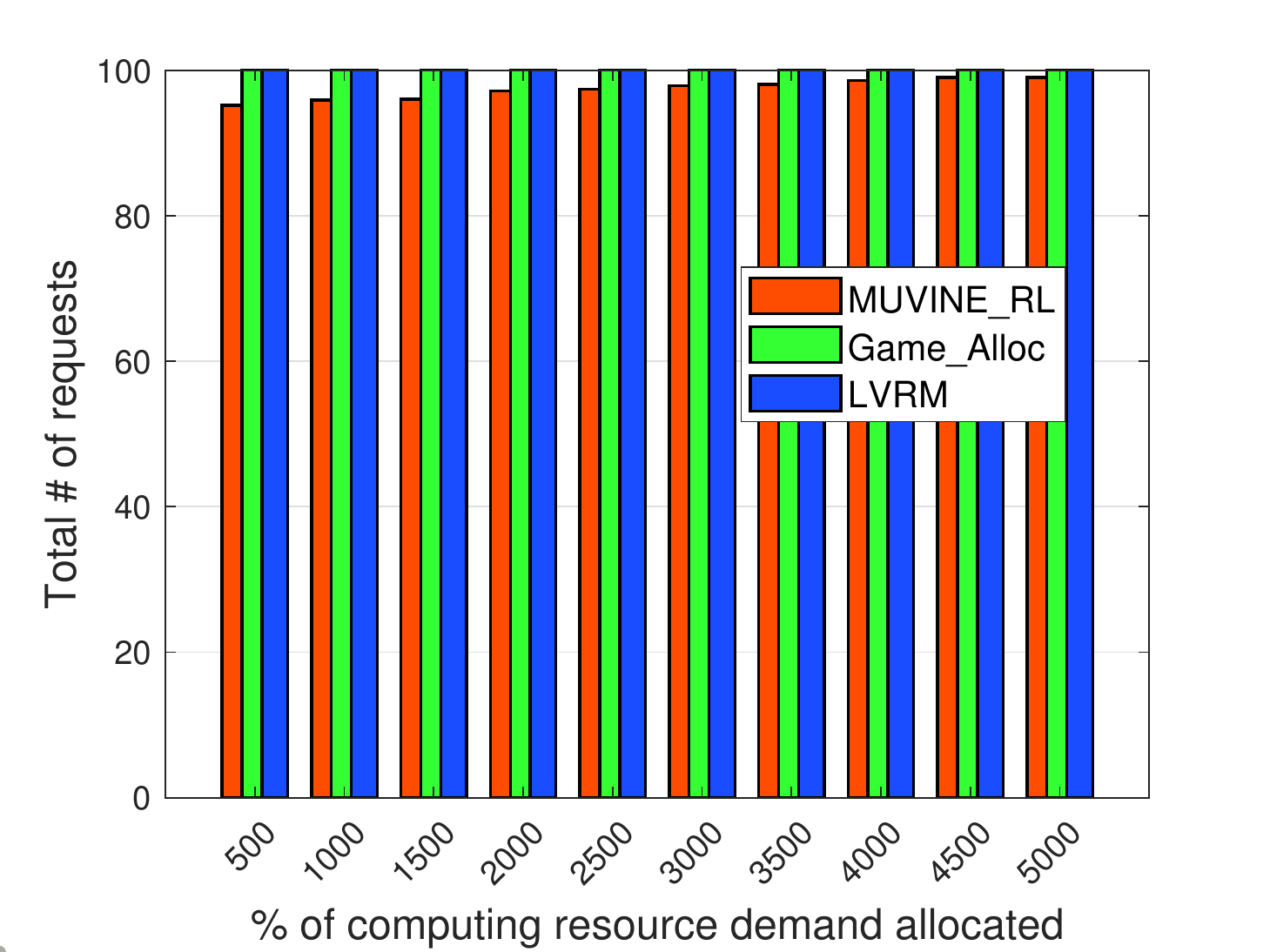}
    \caption{Percentage of computing resource demand allocated}
    \label{fig:AllocatedCompResrc}
\end{figure}

In addition to the time domain evaluation, the VNE should be robust enough to sustain the performance under the varying workload. For that purpose, the existing and proposed schemes are evaluated under the different number of VM requests arrival to know the percentage average CPU and memory utilization. Fig. \ref{fig:3rdstagecpumemoryrurequests}a and Fig. \ref{fig:3rdstagecpumemoryrurequests}b shows the performance comparison of Game\_Alloc \cite{Wei2018Imperfect}, LVRM \cite{sahoo2018lvrm}, Auto\_RP \cite{GHOBAEIARANI2018191}, RL\_VNE \cite{yao2018novel}, and the MUVINE\_RL with average CPU and memory utilization, respectively. The RNN\_VNE \cite{blenk2016boost} scheme is evaluated for only CPU utilization as it does not focus on the memory aspect. It is to note that contrary to the performance under the time domain, with the increase in the number of VM requests arrival, the percentage average resource utilization decreases gradually for each scheme. With compared to the state-of-the-art existing techniques, the MUVINE\_RL improves mean resource utilization by 4.91\%.

The improved performance of MUVINE\_RL over other schemes can be explained as follows. The primary advantage of MUVINE\_RL is its multi-stage prediction model. The improved learning-based admission control of VN requests in the first stage reduces the resource management burden of the CSP and enables the CSP to preserve the resources for the eligible VN requests only. Besides, the maximum likelihood classifier-based VM type prediction in the second stage helps MUVINE\_RL to choose the most suitable SNs and subsequently improves the cloud resource utilization. In the absence of VM classification, the MUVINE\_RL may end up allocating improper substrate resources to VMs, which may further under-utilize or over-utilize the available substrate resources resulting in the inadequate resource utilization. For example, embedding of CPU intensive VMs onto memory intensive SNs and vice versa greatly influences the cloud resource functioning with reduced overall resource utilization. Although, Auto\_RP \cite{GHOBAEIARANI2018191}, RL\_VNE \cite{yao2018novel}, and RNN\_VNE \cite{blenk2016boost} are AI-based schemes, they mainly suffer due to the inclusion of not so important features followed by inadequate information of VM type. This leads to the scenario that the traditional schemes such as Game\_Alloc \cite{Wei2018Imperfect}, LVRM \cite{sahoo2018lvrm} shows the marginal improvement in the performance over the AI-based schemes. The MUVINE\_RL is supported in a three-fold manner from the careful feature selections followed by the prediction of derived features and classification of VMs into respective categories. This leads to the improvement in the overall embedding of VMs and marginally improves the performance in terms of average CPU and memory utilization. 

\begin{figure}
    \centering
    \subfigure[Number of requests vs execution time of proposed scheme.]
    {
        \label{fig:execTime}
        \includegraphics[width=79mm]{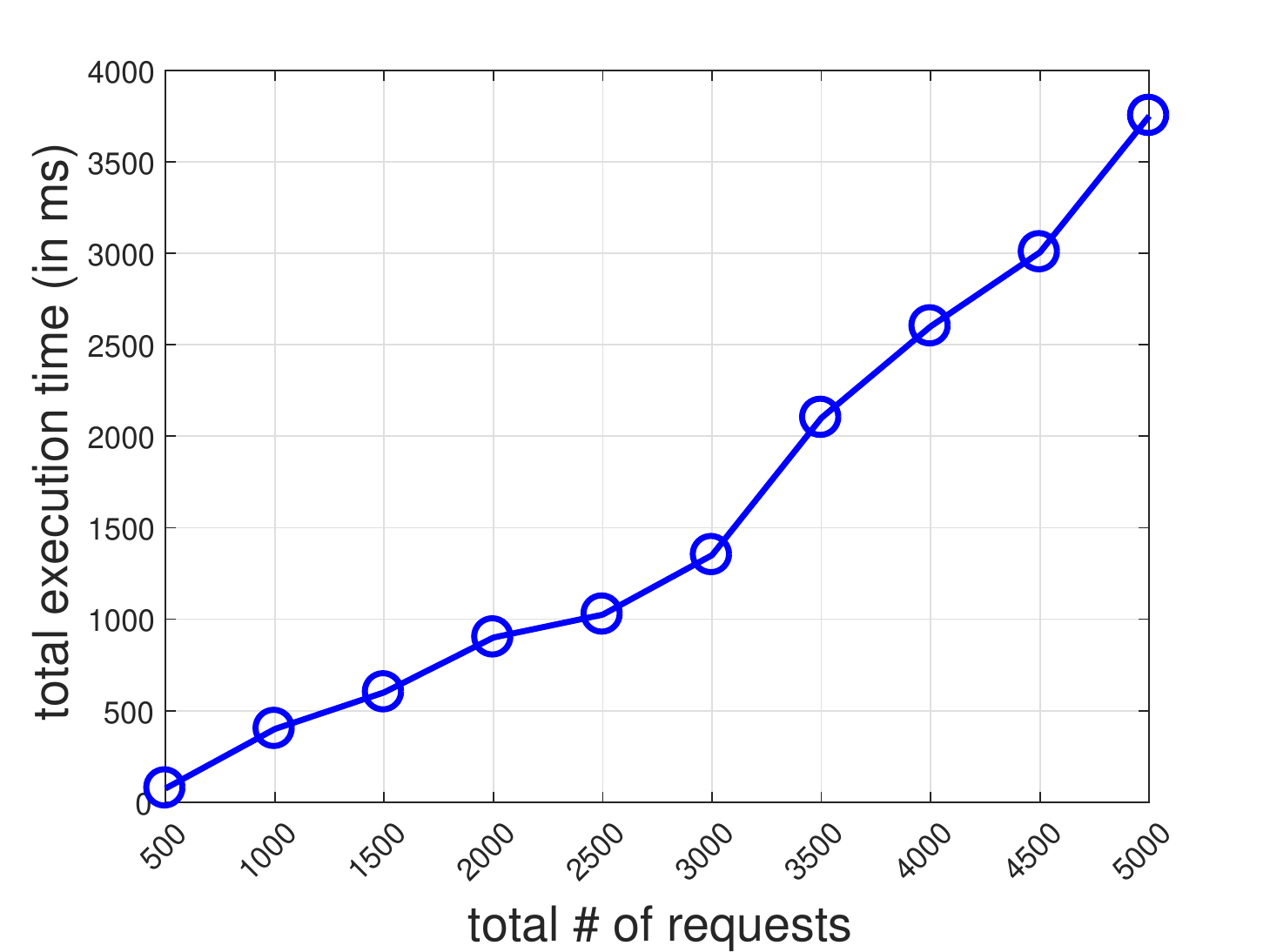}
    }
    \subfigure[Arrival rate vs throughput of proposed scheme.]
    {
        \label{fig:throughput}
        \includegraphics[width=79mm]{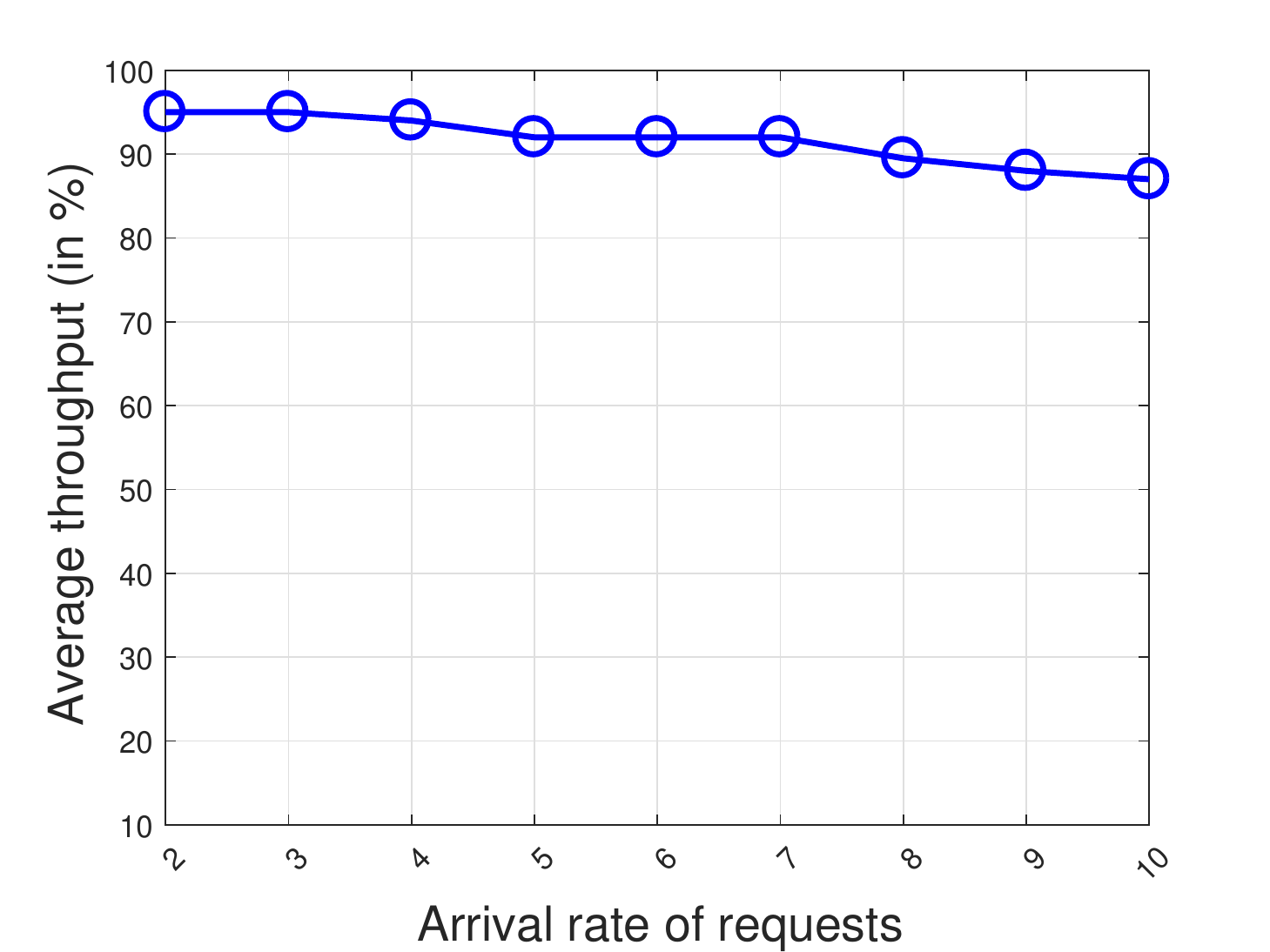}
    }
    \caption{Execution time and throughput of proposed scheme.}
    \label{fig:ExecutionTime_Throughput}
\end{figure}

It is observed that the requests are unable to use cent percent of
the total resource allocated to them. In the proposed MUVINE
scheme, the amount of resource allocated is greater than or equal
to the amount of resource required by the request. As a result,
resource utilization is less, which refers to the ratio of
resource allocated and the amount of resource being used. The
MUVINE\_RL investigate this issue and allocate the resource
according to resource usage prediction. Fig.
\ref{fig:AllocatedCompResrc} shows the comparison result of the
MUVINE\_RL with two recent traditional VNE schemes. The x-axis and
the y-axis in Fig. \ref{fig:AllocatedCompResrc} represent the
number of requests and the percentage of computing resource demand
allocated, respectively. Here requests refer to the virtual
network and computing resource refers to the CPU resource and
memory resource. Both the traditional VNE schemes allocate exactly
the same amount of the resource requested by the VNs. However, the
MUVINE\_RL allocates less amount of resources than the demand. For
$500$ requests a total of $95.3\%$ of the total resource demand is
allocated to the requests. However, as the number of requests
increases to $5000$ a total of $98.8\%$ of the requested resource
are allocated to the VNs.

The performance of the MUVINE\_RL is evaluated in terms of the total execution time and the throughput, as shown in Fig. \ref{fig:execTime} and \ref{fig:throughput}, respectively. The total execution time, which is measured in milliseconds (ms), is observed when the total number of requests increases from $500$ to $5000$. It is observed that the MUVINE\_RL takes total execution time of approximately $75 ms$ to process $500$ requests. However, the execution time increases very rapidly to approximately $400 ms$ when the number of requests is doubled. This rapid increase is due to the randomness of the number of VMs in each request. From the investigation, it is found that the average number of VMs in the first $500$ requests is $3.7$, whereas the average number of VMs in the next $500$ requests only is $7.67$. The total execution time to process $5000$ requests is observed to be $3750 ms$, as shown in Fig. \ref{fig:execTime}. Similarly, to evaluate the performance of the MUVINE\_RL, the throughput is also observed as shown in Fig. \ref{fig:throughput}. The throughput is observed in percentage (in y-axis) with the varied arrival rate of the requests (in x-axis). Throughput refers to the percentage of the total requests that are processed in a single time unit. Arrival rate indicated the number of requests arrived per unit time. The throughput decreases when the arrival increases. The throughput of the MUVINE\_RL is $96\%$ when the arrival rate is $2$ requests per unit time. However, the throughput decreases to approximately $88\%$ when the arrival rate increases to $10$ requests per unit time, as shown in Fig. \ref{fig:throughput}.

\section{Conclusions} \label{sec:conclsn}
In this paper, we deal with the real-time virtual network
embedding problem. Accordingly, Reinforcement Learning based
prediction models are designed for the Multi-stage Virtual Network
Embedding (MUVINE) in cloud data centers. Using the historical
supervised data, the acceptability of real-time incoming VNs is
ascertained using binary supervised classifier followed by
identification of the type of each individual VM. The information
of VM type is used by designing a SARSA reinforcement learning
agent that improves the cloud resource utilization by embedding
the VMs onto the suitable SNs. The binary VN classifier acts as an
admission control and it significantly improves the cloud
performance by rejecting infeasible requests and forwards only
those with higher probability to be accepted. The Radial Basis
Regressor (RBR) model is designed to predict the derived features.
The entire MUVINE scheme is designed in such a way that each
prediction stage contributes towards identifying the appropriate
substrate resources for the given virtual resource demand. Each
stage of the prediction model is extensively simulated and is
evaluated to compare the results with similar state-of-the-art
traditional and AI-based approaches. The simulation results
clearly demonstrate the superiority of our proposed prediction
model over others with consistent outcome across the time domain
and number of virtual requests. However, for further verification
and improvement, we strive to implement the proposed model in the
real cloud environment, which will be part of our future work.

\bibliographystyle{unsrt}
\bibliography{references}

\begin{thebibliography}{10}

\bibitem{zhang2016optimal}
Dong Zhang, Shuhui Li, Min Sun, and Zheng O’Neill.
\newblock An optimal and learning-based demand response and home energy
  management system.
\newblock {\em IEEE Transactions on Smart Grid}, 7(4):1790--1801, 2016.

\bibitem{shalev2016safe}
Shai Shalev-Shwartz, Shaked Shammah, and Amnon Shashua.
\newblock Safe, multi-agent, reinforcement learning for autonomous driving.
\newblock {\em arXiv preprint arXiv:1610.03295}, 2016.

\bibitem{thakkar2019towards}
Hiren~Kumar Thakkar and Prasan~Kumar Sahoo.
\newblock Towards automatic and fast annotation of seismocardiogram signals
  using machine learning.
\newblock {\em IEEE Sensors Journal}, 2019.

\bibitem{mijumbi2015neuro}
Rashid Mijumbi, Juan-Luis Gorricho, Joan Serrat, Meng Shen, Ke~Xu, and Kun
  Yang.
\newblock A neuro-fuzzy approach to self-management of virtual network
  resources.
\newblock {\em Expert systems with applications}, 42(3):1376--1390, 2015.

\bibitem{2018_adaptive}
A.~Alsarhan, A.~Itradat, A.~Y. Al-Dubai, A.~Y. Zomaya, and G.~Min.
\newblock Adaptive resource allocation and provisioning in multi-service cloud
  environments.
\newblock {\em IEEE Transactions on Parallel and Distributed Systems},
  29(1):31--42, Jan 2018.

\bibitem{GHOBAEIARANI2018191}
Mostafa Ghobaei-Arani, Sam Jabbehdari, and Mohammad~Ali Pourmina.
\newblock An autonomic resource provisioning approach for service-based cloud
  applications: A hybrid approach.
\newblock {\em Future Generation Computer Systems}, 78:191 -- 210, 2018.

\bibitem{sahoo2018lvrm}
P.~K. Sahoo, C.~K. Dehury, and B.~Veeravalli.
\newblock Lvrm: On the design of efficient link based virtual resource
  management algorithm for cloud platforms.
\newblock {\em IEEE Transactions on Parallel and Distributed Systems},
  29(4):887--900, April 2018.

\bibitem{Wei2018Imperfect}
W.~Wei, X.~Fan, H.~Song, X.~Fan, and J.~Yang.
\newblock Imperfect information dynamic stackelberg game based resource
  allocation using hidden markov for cloud computing.
\newblock {\em IEEE Transactions on Services Computing}, 11(1):78--89, Jan
  2018.

\bibitem{gao2013multi}
Yongqiang Gao, Haibing Guan, Zhengwei Qi, Yang Hou, and Liang Liu.
\newblock A multi-objective ant colony system algorithm for virtual machine
  placement in cloud computing.
\newblock {\em Journal of Computer and System Sciences}, 79(8):1230--1242,
  2013.

\bibitem{song2019constructive}
An~Song, Wei-Neng Chen, Tianlong Gu, Huaxiang Zhang, and Jun Zhang.
\newblock A constructive particle swarm optimizer for virtual network
  embedding.
\newblock {\em IEEE Transactions on Network Science and Engineering}, 2019.

\bibitem{googleTrace}
Mansaf Alam, Kashish~Ara Shakil, and Shuchi Sethi.
\newblock Analysis and clustering of workload in google cluster trace based on
  resource usage.
\newblock {\em CoRR}, abs/1501.01426, 2015.

\bibitem{xia2017survey}
Wenfeng Xia, Peng Zhao, Yonggang Wen, and Haiyong Xie.
\newblock A survey on data center networking (dcn): Infrastructure and
  operations.
\newblock {\em IEEE communications surveys \& tutorials}, 19(1):640--656, 2017.

\bibitem{2016-5}
Peiying Zhang, Haipeng Yao, and Yunjie Liu.
\newblock Virtual network embedding based on the degree and clustering
  coefficient information.
\newblock {\em {IEEE} Access}, 4:8572--8580, 2016.

\bibitem{pyoung2018joint}
Chan~Kyu Pyoung and Seung~Jun Baek.
\newblock Joint load balancing and energy saving algorithm for virtual network
  embedding in infrastructure providers.
\newblock {\em Computer Communications}, 121:1--18, 2018.

\bibitem{Haeri_2017}
Soroush Haeri and Ljiljana Trajkovic.
\newblock Virtual network embedding via monte carlo tree search.
\newblock {\em {IEEE} Transactions on Cybernetics}, PP(99):1--12, 2017.

\bibitem{2016-VNE}
F.~Esposito, I.~Matta, and Y.~Wang.
\newblock Vinea: An architecture for virtual network embedding policy
  programmability.
\newblock {\em IEEE Transactions on Parallel and Distributed Systems},
  27(11):3381--3396, Nov 2016.

\bibitem{8314665}
N.~Shahriar, S.~R. Chowdhury, R.~Ahmed, A.~Khan, S.~Fathi, R.~Boutaba,
  J.~Mitra, and L.~Liu.
\newblock Virtual network survivability through joint spare capacity allocation
  and embedding.
\newblock {\em IEEE Journal on Selected Areas in Communications}, pages 1--1,
  2018.

\bibitem{zhang2016resource}
Jiangtao Zhang, Hejiao Huang, and Xuan Wang.
\newblock Resource provision algorithms in cloud computing: A survey.
\newblock {\em Journal of Network and Computer Applications}, 64:23--42, 2016.

\bibitem{Aral201689}
Atakan Aral and Tolga Ovatman.
\newblock Network-aware embedding of virtual machine clusters onto federated
  cloud infrastructure.
\newblock {\em Journal of Systems and Software}, 120:89 -- 104, 2016.

\bibitem{2016-3}
Long Gong, Huihui Jiang, Yixiang Wang, and Zuqing Zhu.
\newblock Novel location-constrained virtual network embedding {LC}-{VNE}
  algorithms towards integrated node and link mapping.
\newblock {\em {IEEE}/{ACM} Transactions on Networking}, 24(6):3648--3661, dec
  2016.

\bibitem{2016-7-Yin}
Lei Yin, Zheng Chen, Ling Qiu, and Yonggang Wen.
\newblock Interference based virtual network embedding.
\newblock In {\em 2016 {IEEE} International Conference on Communications
  ({ICC})}. {IEEE}, may 2016.

\bibitem{song2019divide}
An~Song, Wei-Neng Chen, Yue-Jiao Gong, Xiaonan Luo, and Jun Zhang.
\newblock A divide-and-conquer evolutionary algorithm for large-scale virtual
  network embedding.
\newblock {\em IEEE Transactions on Evolutionary Computation}, 2019.

\bibitem{song2019distributed}
An~Song, Wei-Neng Chen, Tianlong Gu, Huaqiang Yuan, Sam Kwong, and Jun Zhang.
\newblock Distributed virtual network embedding system with historical archives
  and set-based particle swarm optimization.
\newblock {\em IEEE Transactions on Systems, Man, and Cybernetics: Systems},
  2019.

\bibitem{2018_reinforcement_VNE}
Haipeng Yao, Xu~Chen, Maozhen Li, Peiying Zhang, and Luyao Wang.
\newblock A novel reinforcement learning algorithm for virtual network
  embedding.
\newblock {\em Neurocomputing}, 284:1 -- 9, 2018.

\bibitem{zhang2018intelligent}
Yu~Zhang, Jianguo Yao, and Haibing Guan.
\newblock Intelligent cloud resource management with deep reinforcement
  learning.
\newblock {\em IEEE Cloud Computing}, 4(6):60--69, 2018.

\bibitem{wang2019coordinated}
Cong Wang, Fanghui Zheng, Sancheng Peng, Zejie Tian, Yujia Guo, and Ying Yuan.
\newblock A coordinated two-stages virtual network embedding algorithm based on
  reinforcement learning.
\newblock In {\em 2019 Seventh International Conference on Advanced Cloud and
  Big Data (CBD)}, pages 43--48. IEEE, 2019.

\bibitem{yao2018rdam}
Haipeng Yao, Bo~Zhang, Peiying Zhang, Sheng Wu, Chunxiao Jiang, and Song Guo.
\newblock Rdam: A reinforcement learning based dynamic attribute matrix
  representation for virtual network embedding.
\newblock {\em IEEE Transactions on Emerging Topics in Computing}, 2018.

\bibitem{zhang2018efficient}
Q.~Zhang, L.~T. Yang, Z.~Yan, Z.~Chen, and P.~Li.
\newblock An efficient deep learning model to predict cloud workload for
  industry informatics.
\newblock {\em IEEE Transactions on Industrial Informatics}, 14(7):3170--3178,
  July 2018.

\bibitem{sagnika2018workflow}
Santwana Sagnika, Saurabh Bilgaiyan, and Bhabani Shankar~Prasad Mishra.
\newblock Workflow scheduling in cloud computing environment using bat
  algorithm.
\newblock In {\em Proceedings of First International Conference on Smart
  System, Innovations and Computing}, pages 149--163. Springer, 2018.

\bibitem{blenk2016boost}
Andreas Blenk, Patrick Kalmbach, Patrick Van Der~Smagt, and Wolfgang Kellerer.
\newblock Boost online virtual network embedding: Using neural networks for
  admission control.
\newblock In {\em Network and Service Management (CNSM), 2016 12th
  International Conference on}, pages 10--18. IEEE, 2016.

\bibitem{dehury2019dyvine}
Chinmaya~Kumar Dehury and Prasan~Kumar Sahoo.
\newblock Dyvine: Fitness-based dynamic virtual network embedding in cloud
  computing.
\newblock {\em IEEE Journal on Selected Areas in Communications},
  37(5):1029--1045, 2019.

\bibitem{scikit-learn}
F.~Pedregosa, G.~Varoquaux, A.~Gramfort, V.~Michel, B.~Thirion, O.~Grisel,
  M.~Blondel, P.~Prettenhofer, R.~Weiss, V.~Dubourg, J.~Vanderplas, A.~Passos,
  D.~Cournapeau, M.~Brucher, M.~Perrot, and E.~Duchesnay.
\newblock Scikit-learn: Machine learning in {P}ython.
\newblock {\em Journal of Machine Learning Research}, 12:2825--2830, 2011.

\bibitem{abadi1603tensorflow}
M~Abadi, A~Agarwal, P~Barham, E~Brevdo, Z~Chen, C~Citro, GS~Corrado, A~Davis,
  J~Dean, M~Devin, et~al.
\newblock Tensorflow: large-scale machine learning on heterogeneous distributed
  systems. arxiv preprint (2016).
\newblock {\em arXiv preprint arXiv:1603.04467}.

\bibitem{yao2018novel}
Haipeng Yao, Xu~Chen, Maozhen Li, Peiying Zhang, and Luyao Wang.
\newblock A novel reinforcement learning algorithm for virtual network
  embedding.
\newblock {\em Neurocomputing}, 284:1--9, 2018.

\end{thebibliography}

\end{document}